\documentclass[10pt]{article}

\usepackage{amsmath,amssymb,amsthm}
\usepackage{amsxtra}
\usepackage{amsfonts}
\usepackage{amssymb}
\usepackage{url}
\usepackage{hyperref}
\usepackage{esint}
\usepackage{eucal}%
\usepackage{mathrsfs}
\usepackage{ucs}
\usepackage[utf8x]{inputenc}
\usepackage{cite}
\usepackage{graphicx,graphics}
\usepackage{subcaption}
\newcommand\fig[1] {{\rm Figure}~\ref{fig:#1}}
\newcommand\labfig[1] {\label{fig:#1}}

\newcommand\labsect[1] {\label{sect:#1}}

\newcommand\eq[1] {(\ref{#1})}
\newcommand{\bfm}[1]{\mbox{\boldmath ${#1}$}}

\newcommand{\nonum}{\nonumber \\}
\newcommand{\beqa}{\begin{eqnarray}}
\newcommand{\eeqa}[1]{\label{#1}\end{eqnarray}}
\newcommand{\beq}{\begin{equation}}
  \newcommand{\eeq}[1]{\label{#1}\end{equation}}

\newcommand{\Grad}{\nabla}
\newcommand{\Div}{\nabla \cdot}

\newcommand{\Md}{\partial}

\newcommand{\Ga}{\alpha}
\newcommand{\Gb}{\beta}
\newcommand{\Gd}{\delta}

\newcommand{\Gf}{\phi}

\newcommand{\Gg}{\gamma}

\newcommand{\Gk}{\kappa}

\newcommand{\Gl}{\lambda}

\newcommand{\Gm}{\mu}

\newcommand{\Gt}{\theta}

\newcommand{\Gvr}{\varrho}
\newcommand{\Gs}{\sigma}

\newcommand{\GG}{\Gamma}
\newcommand{\GL}{\Lambda}

\newcommand{\BGs}{\bfm\sigma}

\newcommand{\bpm}{\begin{pmatrix}}
\newcommand{\epm}{\end{pmatrix}}

\def\Bk{{\bf k}}

\def\Bn{{\bf n}}

\def\Br{{\bf r}}

\def\Bx{{\bf x}}

\def\Bz{{\bf z}}
\def\BA{{\bf A}}

\def\BI{{\bf I}}

\def\BN{{\bf N}}

\def\BT{{\bf T}}
\def\BU{{\bf U}}
\def\BV{{\bf V}}

\usepackage{graphics}
\usepackage{amssymb}
\def\presuper#1#2%
  {\mathop{}%
   \mathopen{\vphantom{#2}}^{#1}%
   \kern-\scriptspace%
   #2}

 \title{A possible explanation of dark matter and dark energy involving a vector torsion field}
\author{Graeme W. Milton \footnote{Department of Mathematics, University of Utah, USA -- milton@math.utah.edu,}}
\date{}
\begin{document}
\maketitle
\vskip -.5cm 
\vskip 1.cm
\begin{abstract}
  A simple gravitational model with torsion is studied, and it is suggested that it could explain the dark
  matter and dark energy in the universe. It can be reinterpreted as a model using the Einstein gravitational
  equations where spacetime has regions filled with a perfect fluid with negative energy (pressure) and
  positive mass density, other regions containing an anisotropic substance that in the rest frame (where the momentum is zero)
  has negative mass density and a uniaxial stress tensor, and possibly other ``luminal'' regions where there is no rest
  frame. The torsion vector field is inhomogeneous throughout spacetime, and possibly turbulent. Numerical simulations
  should reveal whether or not the equations are consistent with cosmological observations of dark matter and dark energy. 
\end{abstract}

\section{Introduction}
\setcounter{equation}{0}
One of the outstanding problems in physics is to account for the apparent dark energy and dark matter in the
universe since it accounts for roughly $95\%$ of total matter in the universe.
Reviews of the dark matter and dark energy cosmological problem, and the models that have been introduced to
account for it, include those of Peebles and Ratra \cite{Peebles:2003:CCD}, Sahni \cite{Sahni:2004:CAI}, 
Copeland, Sami, and Tsujikawa \cite{Copeland:2006:DDE},
Frieman, Turner, and Huterer \cite{Frieman:2008:DEA}, Amendola and Tsujikawa \cite{Amendola:2010:DET},
Li, Li, Wang and Wang \cite{Li:2016:DE},
and Arun, Gudennavar and Sivaram \cite{Arun:2017:DMD}.
We will not survey the literature here as these reviews do an excellent job of that. As is often the case, we use dimensions where the
speed of light $c$ is $1$, we use the Einstein summation convention where sums over repeated indices are assumed, and a comma
in front of a lower index such as $f_{,i}$ denotes differentiation of $f$ with respect to $x^i$.

Maybe the most favored model is the $\GL$CDM model. Here $\GL$ is Einstein's cosmological constant,
giving rise to dark energy with
$p=-\Gm_0$ and CDM is cold dark matter introduced to give the observed ratio of pressure to total mass density
which is about $-0.8$. Constraints on dark matter and dark energy properties are imposed by results of the DES collaboration
\cite{Abbott:2019:CCM, Nadler:2021:MWS}. Gravitational-lensing measurements \cite{Wong:2020:HMH} give a Hubble constant
that is consistent with long period Cepheid measurements in the large Magellanic cloud \cite{Reiss:2019:LMC} 
but both strongly indicate significant discrepancies with the  $\GL$CDM model. Experimental tests of the
strong equivalence principle \cite{Chae:2020:TSE} provide further evidence casting doubt on the model in favor of
modified gravity theories.

The relativistic model we introduce here has no adjustable parameters and incorporates a torsion vector field.
It is perhaps the simplest gravitational model involving torsion, yet we believe it may explain the dark energy
and dark mass in the universe. If simplicity of the underlying equations is to be a guiding principle in physics,
then these equations surely meet that principle. Of course, our equations still need be compatible with both existing
and future experimental observations, both qualitatively and quantitatively, and this remains to be seen. It is to be
emphasized that our equations govern the curvature of empty space and do not fully determine the interaction between matter
and the curvature. We believe the simpler problem of obtaining the equations
for empty space should be addressed first, as a stepping stone towards a more general theory where matter is included.
The main demands that drive our formulation of the equations are:
\begin{itemize}
\item That the new equations should be as simple as possible, involving as few assumptions as possible
\item That, correspondingly, the new equations should be linear constraints on the curvature tensor.
\item That, clearly, the number of unknowns in the torsion field and in the metric, modulo coordinate transformations,
  should be equal to the number of independent scalar constraints imposed by the new equations.
\item That any solution to Einstein's equations be also a solution to the new equations.
\end{itemize}
It may be argued that these should not be assumed apriori, but that convincing physical arguments should be presented
as well. On the other hand, Einstein's equations for empty space can be obtained from the first three of these requirements
without any necessity to introduce physical considerations. Only when matter is present is physics needed to determine the full Einstein equations,
as embodied in the constraints that the equations
reduce to Newton's gravitational equations when the space-time curvature is small and that small test particles follow geodesics. 
Since we do not consider the full interaction
of matter and curvature we cannot claim that small test particles will still follow geodesics: that would be
a natural demand to be required of a more general theory.

Despite the simplicity of our underlying equations
the resultant dynamics of the torsion vector field, even in the weak field approximation, is enormously complicated,
suggesting the torsion vector field has some sort of turbulent behavior. This is the main novel feature of our theory: the suggestion
that space itself is intrinsically inhomogeneous on many length scales,  even in the absence of matter. This goes further
than the idea that space is inhomogeneous on the Planck length scale. 

Numerical simulations of the torsion field behavior
will almost certainly be necessary to test the theory and assess its compatibility with astronomical and cosmological
observations. The equations can be reinterpreted as a model using the Einstein gravitational
equations where spacetime has regions
filled with a perfect fluid with negative energy (pressure) and positive mass density, other regions containing an
anisotropic substance that in the local rest frame (where the momentum is zero) has negative mass density and a uniaxial stress tensor,
and possibly other ``luminal'' regions where there is no natural local ``rest frame''. We emphasize, though,
that all three regions are manifestations of the torsion vector field, and the three regions accordingly correspond to
regions where the vector field points inside, outside, or on the boundary of the light cone.  It has been noted before
by De Sabbata and Sivaram \cite{Sabbata:1994:STG} that torsion provides a natural framework for negative mass, as has
been suggested to occur in the early universe. Cosmological models with negative mass have been studied by
Ray, Khlopov, Ghosh and Mukhopadhyay \cite{Ray:2011:PLM} and by
Famaey and McGaugh \cite{Famaey:2012:MND} and yield promising explanations for the acceleration of the expansion rate of the universe.

In our theory dark
energy and dark matter interact. Other models where dark energy and dark matter interact are reviewed by
Wang, Abdalla, Atrio-Barandela and Pav\'on \cite{Wang:2016:DMD}. 

In additional to the cosmological dark mass problem there is also the
dark mass problem that is associated with the observations
of higher than expected rotational velocities of stars far from the galactic center. One empirically motivated
model that successfully accounts for
this is MOND (Modified Newtonian Dynamics), first introduced by Milgrom \cite{Milgrom:1983:MNDIII}. He suggested that
Newton's law, where the gravitational force is proportional to the acceleration be replaced at low accelerations, below a critical
acceleration $a_0=\approx 1.2\times 10^{-10}ms^{-2}$, by one where the
force is proportional to the square of the acceleration, see \fig{moti}.
Later this idea motivated a relativistic theory
developed by Bekenstein \cite{Bekenstein:2004:RGT} and generalized by Skordis \cite{Skordis:2008:GTV}. One prediction of MOND,
later verified, was that there should be a universal relation between between the rotation speeds of stars in the outermost parts
of a galaxy and the total mass, not dark mass, of the galaxy: see the book of Merritt \cite{Merritt:2020:PAM} for further discussion
on this point. In particular, on the basis of this, it seems unlikely that unseen particles will provide the explanation for
the galactic missing mass problem. 
Other reviews of MOND, including these and other relativistic extensions and their implications for cosmology,
have been given by Famaey and McGaugh \cite{Famaey:2012:MND}, Merritt \cite{Merritt:2020:PAM} and Milgrom \cite{Milgrom:2020:MON}.
It is not yet clear whether the torsion field model developed here will be successful in explaining the galactic
dark mass problem, though the success of Farnes \cite{Farnes:2018:UTD} in explaining the flattening of rotation curves by
introducing negative mass suggests that it might meet with success on this front.

\begin{figure}[!ht]
\includegraphics[width=0.9\textwidth]{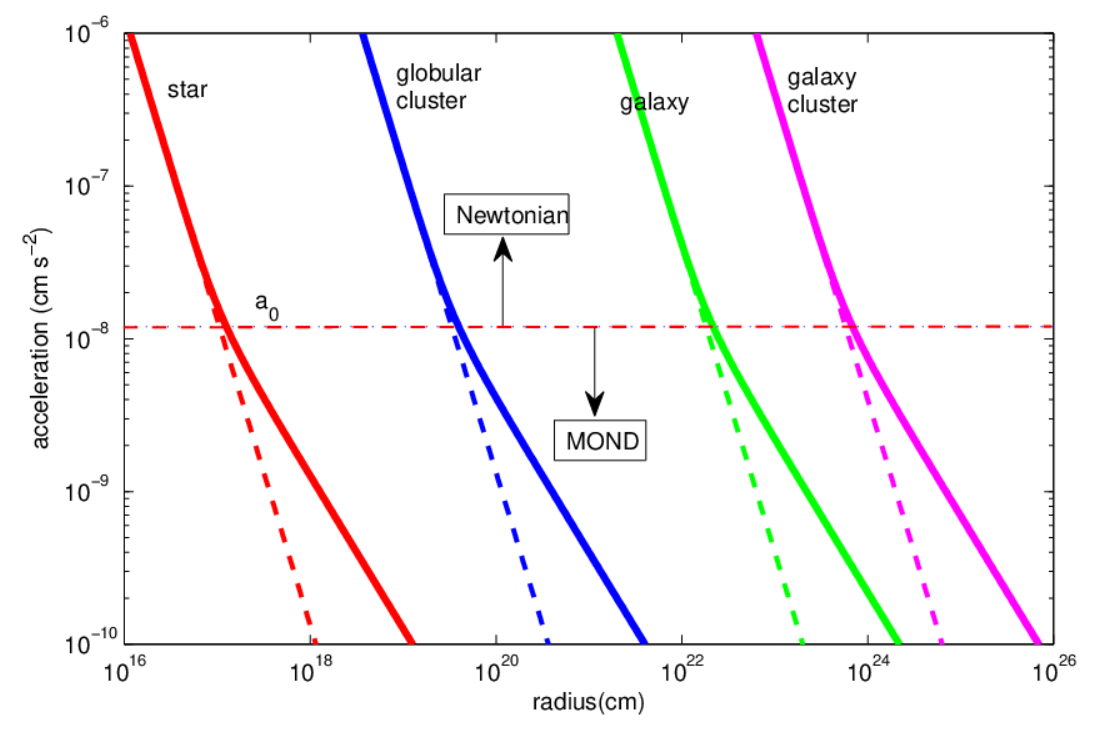}
\vskip -0.0cm
\caption{Figure, courtesy of M. Milgrom, taken from \protect{\url{http://www.scholarpedia.org/article/The_MOND_paradigm_of_modified_dynamics}}
  showing its predictions, that are consistent with experimental observations. Plotted is the acceleration as a function of the distance from an isolated mass $M$,
  for a star with $M=M_\odot$ (red), a globular cluster with $M=10^5M_\odot$ (blue), a galaxy with $M=3\times 10^{10}M_\odot$ (green),
  and a galaxy cluster with $M=3\times 10^{13}M_\odot$ (magenta), in which $ M_\odot$ represents one solar mass.}
  \labfig{moti}
\end{figure}

Torsion is the antisymmetric part of the affine connection. The affine connection determines how vectors change under
parallel displacements. Cartan introduced torsion and applied it to develop generalizations of Einstein's
gravitational equations. His work dates back to the early 1920's:
see \cite{Cartan:1955:MAC} and references therein (translated in \cite{Magnon:1986:MAC}).
A brief introduction to torsion is in the classic book on gravitation by Misner, Thorne, and Wheeler \cite{Misner:1973:G}.
More extensive reviews of general relativistic models that include torsion, with further developments, include those of
Hehl, von der Heyde, and Kerlick \cite{Hehl:1976:GRS}, De Sabbata and Sivaram \cite{Sabbata:1994:STG},
Hehl, McCrea, Mielke and Ne'eman \cite{Hehl:1995:MAG},
Shapiro \cite{Shapiro:2002:PAS}, and Poplawski \cite{Poplawski:2009:SF}. Interestingly, Jose Beltr{\'a}n Jim{\'e}neza,
Lavinia Heisenberg and Tomi S. Koivisto have recently shown \cite{Beltran:2019:GTG} that Einstein's gravitational
equations can be reformulated in terms of the torsion alone, eliminating the metric.  

Typically, general relativistic models with torsion
have been introduced to allow for the intrinsic spin of matter, and are quite complicated.
By contrast, our focus here is on developing
a simple model that may account for the dark mass and energy in the universe. 

Ivanov and Wellenzohn have suggested that the Einstein-Cartan theory may account for Dark Energy \cite{Ivanov:2016:ECG}.
Another gravitational model that incorporates the same torsion vector field we use, as well as additional fields and
a fifth dimension, has been developed by Sengupta \cite{Sengupta:2020:GTD} who suggests it may solve both the cosmological
and galactic dark matter problem. Other models incorporating torsion, quite different to the one explored here, that
may explain the accelerated expansion of the universe have been developed by Watanabe and Hayashi \cite{Watanabe:2004:GRT},
Minkevich \cite{Minkevich:2009:AUS}, de Berredo-Peixoto and de Freitas \cite{Berredo-Peixoto:2009:TER}, 
Belyaev, Thomas and Shapiro \cite{Belyaev:2017:TDM}, and  Vasak, Kirsch, and Struckmeier \cite{Vasak:2020:DEI}.

The analysis in the following sections is more or less standard, though equivalent formulations are clearly possible
according to one's mathematical taste. The key step to arriving at our equations is simply to postulate that
geodesics and autoparallels coincide. There is nothing difficult in the analysis leading to our
equations governing the spacetime curvature.


\section{Metric and Affinities}
\setcounter{equation}{0}
The functions $g_{uv}$ of the metric field describe with respect to the
arbitrarily chosen system of co-ordinates the metrical relations of the
spacetime continuum:

\beq ds^2=g_{uv}dx^udx^v.
\eeq{1.1}
Here we will assume that the $g_{uv}$ are real and symmetric in the
indices $u$ and $v$
and thus \eq{1.1} provides the defining equation for $g_{uv}$ with respect
to a given coordinate system.

Now consider the affinity $\Gamma^{i}_{st}$ which determines a vector
after parallel displacement. To a real contravariant vector $\BA$
with components $A^i$ at a point $P$ with coordinates $x^t$, we correlate
a vector $\BA+\Gd\BA$ with components $A^i+\Gd A^i$ at the infinitesimally
close point with coordinates $x^t+\Gd x^t$ by

\beq \Gd A^i=-\GG^i_{st}A^s \Gd x^t. \eeq{1.2}

Since the magnitude of $\BA$ in parallel displacement does not change to first order in that displacement
we obtain
\beq 0=\Gd[g_{uv}A^u A^v]=\frac{d g_{uv}}{dx^\Ga}A^uA^v dx^\Ga
+g_{uv}A^u(\Gd A^v) +g_{uv}A^v(\Gd A^u),
\eeq{1.3}
and so, using \eq{1.2}, we get
\beq g_{uv,\Ga}-g_{u\Gb}\GG^\Gb_{v\Ga}-g_{v\Gb}\GG^\Gb_{u  \Ga}=0,
\eeq{1.4}
where the comma denotes partial differentiation. Now by considering this equation together with the two
equations
\beq g_{v\Ga,u}-g_{v\Gb}\GG^\Gb_{\Ga u}-g_{\Ga\Gb}\GG^\Gb_{v u}=0,
\eeq{1.5}
\beq g_{\Ga u,v}-g_{\Ga\Gb}\GG^\Gb_{u v}-g_{u\Gb}\GG^\Gb_{\Ga v}=0,
\eeq{1.6}
that obtained are obtained by a cyclic interchange of indices,
and by subtracting \eq{1.4} from the sum of \eq{1.5} and \eq{1.6} 
we get
\beq [u~v,\Ga]+g_{v\Gb}{\widehat{\GG}}^\Gb_{u\Ga}
+g_{u\Gb}\widehat\GG^\Gb_{v\Ga}-g_{\Ga\Gb}{\widehat{\GG}}^\Gb_{uv}=
g_{\Ga\Gb}\GG^\Gb_{uv},
\eeq{1.7}
where $[u~v,\Ga]$ is the Christoffel symbol of the first kind, given by
\beq [u~v,\Ga]=\tfrac{1}{2}(g_{\Ga u,v}+g_{\Ga v,u}-g_{uv,\Ga}),
\quad \widehat{\GG}^\Gb_{ij}=\tfrac{1}{2}(\GG^\Gb_{ij}-\GG^\Gb_{ji}).
\eeq{1.8}
The antisymmetric part of the affinity $\widehat{\GG}^\Gb_{ij}$, in contrast to  $\GG^\Gb_{ji}$, is a tensor - Cartan's torsion tensor.  
\section{Equating Geodesics with Autoparallels}
\setcounter{equation}{0}
Geodesics are trajectories $\Bx(s)$, which we choose to parametrize by
the distance $s$ along them, that have an extremal distance between two points.
Since they clearly only depend on the metric they satisfy the standard formula:
\beq \frac{d^2x^\Gm}{ds^2}
+g^{\Gm r}[\Ga~\Gb,r]\frac{dx^\Ga}{ds}\frac{dx^\Gb}{ds}=0.
\eeq{1.9}
Alternatively we may consider an autoparallel constructed in such a way that
successive elements arise from each other by parallel displacements.
An element is the vector $d\Bx/ds$ and under parallel displacement its
components transform as
\beq \Gd\left(\frac{dx^u}{ds}\right)=-\GG^\Gm_{\Ga\Gb}\frac{dx^\Ga}{ds}\Gd x^\Gb.
\eeq{1.10}
The left hand side is to be replaced by $(d^2x^\Gm/ds^2)\Gd s$ giving
\beq \frac{d^2x^\Gm}{ds^2}+\GG^\Gm_{\Ga\Gb}\frac{dx^\Ga}{ds}\frac{dx^\Gb}{ds}.
\eeq{1.11}
We postulate that geodesics coincide with autoparallels, thus giving
\beq \left\{\GG^\Gm_{\Ga\Gb}-
  g^{\Gm r} [\Ga~\Gb,r]\right\}\frac{dx^\Ga}{ds}\frac{dx^\Gb}{ds}=0,
\eeq{1.12}
or equivalently
\beq \left\{\tfrac{1}{2}(\GG^\Gm_{\Ga\Gb}+\GG^\Gm_{\Gb\Ga})-
g^{\Gm r}[\Ga~\Gb,r]\right\}\frac{dx^\Ga}{ds}\frac{dx^\Gb}{ds}=0.
\eeq{1.13}
This postulate is fundamental to the theory. While it is absent of any physical
justification, aside from removing possible ambiguity in the path that test particles
are required to follow in a more general theory, it is essential to keep the governing equations
as simple as possible. This is our motivation for this constraint. 

As \eq{1.13} holds for all $dx^\Ga/ds$ and  $dx^\Gb/ds$ we obtain
\beq \overline{\GG}^\Gm_{\Ga\Gb}\equiv\tfrac{1}{2}(\GG^\Gm_{\Ga\Gb}+\GG^\Gm_{\Gb\Ga})=
g^{\Gm r} [\Ga~\Gb,r].
\eeq{1.14}
Multiplying both sides by $g_{\Gm s}$ and summing over $\Gm$ gives
\beq g_{\Gm s}\GG^\Gm_{\Ga\Gb}+g_{\Gm s}\GG^\Gm_{\Gb\Ga}=2[\Ga~\Gb, s].
\eeq{1.15}
Combining this with \eq{1.7} then yields
\beq S_{\Ga\Gb\Gm}\equiv g_{\Ga r}\widehat\GG^r_{\Gb\Gm}
=-g_{\Gb r}\widehat\GG^r_{\Ga\Gm}=-S_{\Gb\Ga\Gm}=S_{\Gb\Gm\Ga}.
\eeq{1.16}
So $S_{\Ga\Gb\Gm}$ is antisymmetric with respect to interchange of any pair
of its three indices and this implies (see, for example, the text below equation (2.16) in \cite{Hehl:1976:GRS}) that 
\beq \widehat\GG^i_{jk}=g^{ir}e_{rjk\ell}U^{\ell}, \eeq{1.17}
for some contravariant vector density $\BU$ where, as standard, $e_{rjk\ell}$
is the Levi-Civita tensor density, with
$e_{1234}=1$ and which antisymmetric with respect to interchange of any
pair of indices. $\BU$ is known as the axial part of the torsion \cite{Hehl:1976:GRS}.  Combining \eq{1.17} with \eq{1.14} gives
\beq \GG^\Gm_{\Ga\Gb}=\overline{\GG}^\Gm_{\Ga\Gb}+g^{\Gm r}e_{r\Ga\Gb\ell}U^{\ell}.
\eeq{1.18}

\section{The Ricci Tensor}
\setcounter{equation}{0}
Let us express the Ricci Tensor
\beq R_{jk}  =  \GG^i_{ri}\GG^r_{jk}-\GG^i_{rk}\GG^r_{ji}
+\GG^i_{jk,i}-\GG^i_{ji,k},
\eeq{1.21}
that is associated with the local curvature of spacetime, in terms of the symmetric and antisymmetric parts of the affinity:
\beqa
R_{jk}&= &(\overline{\GG}^i_{ri}+\widehat\GG^i_{ri})(\overline{\GG}^r_{jk}+\widehat\GG^r_{jk}) -(\overline{\GG}^i_{rk}+\widehat\GG^i_{rk})(\overline{\GG}^r_{ji}+\widehat\GG^r_{ji})
+(\overline{\GG}^i_{jk}+\widehat\GG^i_{jk})_{,i}
-(\overline{\GG}^i_{ji}+\widehat\GG^i_{ji})_{,k}
,\nonum
&= &\overline{\GG}^i_{ri}(\overline{\GG}^r_{jk}+\widehat\GG^r_{jk})
-(\overline{\GG}^i_{rk}+\widehat\GG^i_{rk})(\overline{\GG}^r_{ji}+\widehat\GG^r_{ji})
+(\overline{\GG}^i_{jk}+\widehat\GG^i_{jk})_{,i}
-\overline{\GG}^i_{ji,k},
\eeqa{1.22}
where we have used the fact that $\GG^i_{ri}=0$ as follows from \eq{1.17}.
So now we have
\beq R_{jk}=R^0_{jk}-\widehat\GG^i_{rk}\widehat\GG^r_{ji}
+\widehat\GG^i_{kr}\overline{\GG}^r_{ji}+\overline{\GG}^i_{rk}\widehat\GG^r_{ij}
-\overline{\GG}^i_{ri}\widehat\GG^r_{kj}
-\widehat\GG^i_{kj,i},
\eeq{1.23}
where
\beq R^0_{jk}=\overline{\GG}^i_{ri} \overline{\GG}^r_{jk}- \overline{\GG}^i_{rk} \overline{\GG}^r_{ji}+\overline{\GG}^i_{jk,i}
-\overline{\GG}^i_{ji,k}
\eeq{1.24}
is the usual Ricci curvature tensor associated just with the metric. We now consider the symmetric part of $R_{jk}$ as it is
central to our equations:
\beq \overline{R}_{jk}\equiv\tfrac{1}{2}(R_{jk}+R_{kj})
=R^0_{jk}-\widehat\GG^i_{rk}\widehat\GG^r_{ji}
=R^0_{jk}-g^{si}e_{srkl}U^\ell g^{tr}e_{tjih}U^h.
\eeq{1.25}
Given an arbitrary point we can always find a new coordinate system such that the metric is orthogonal at that point. In this new coordinate system at this one
point
\beqa \overline{R}_{11} & = &
R^0_{11}-g^{ii}e_{ir1\ell}U^\ell g^{rr}e_{r1ih}U^h ,\nonum
\overline{R}_{12} & = & R^0_{11}-g^{ii}e_{ir1\ell}U^\ell g^{rr}e_{r2ih}U^h,
\eeqa{1.26}
where a sum over $i$ and $r$ is implied. For $e_{ir1\ell}e_{r1ih}$ to be non-zero, it is necessary that
$r\ne i$ and $ir\ell$ must be a permutation of $rih$
(and a permutation of $234$), implying $\ell=h$. So we obtain
\beq \overline{R}_{11}=R^0_{11}-2g^{22}g^{33}(U^4)^2
-2g^{44}g^{22}(U^3)^2-2g^{33}g^{44}(U^2)^2.
\eeq{1.27}
Also for $e_{ir1\ell}e_{r2ih}$ to be nonzero $\ell$ must be 2 and $h$ must be $1$,
implying
\beq \overline{R}_{12}=R^0_{12}+2g^{33}g^{44}U^3U^4. \eeq{1.28}
Of course, similar formulas hold for the other elements of $\overline{R}_{jk}$.
Hence at this point, in this coordinate system,
\beq \overline{R}_{jk}
=R^0_{jk}+2g^{-1}g_{jn}g_{km}U^mU^n-2g^{-1}g_{jk}g_{mn}U^mU^n,
\eeq{1.29}
where $g=g_{11}g_{22}g_{33}g_{44}$ is the determinant of the metric tensor.
Or, introducing a contravariant vector $N^k$ such that $N^k=U^k/\sqrt{-g}$
we obtain
\beq \overline{R}_{jk}
=R^0_{jk}+2g_{jk}g_{mn}N^mN^n-2g_{jn}g_{km}N^mN^n.
\eeq{1.30}
This equation being a tensor equation will be true in any coordinate system,
and also at ant point since the original point was arbitrarily chosen.
Raising indices gives
\beq  \overline{R}^j_{k}=(R^0)^j_{k}+2\Gd^j_k g_{mn}N^mN^n-2g_{km}N^mN^j.
\eeq{1.31}
Finally, contracting indices we get
\beq \overline{R}\equiv\overline{R}^j_{j}=R^0+6g_{mn}N^mN^n,
\eeq{1.32}
where $R^0=(R^0)^j_{j}$. We will call $\BN$ the torsion field. 
\section{The proposed new gravitational equations}
\setcounter{equation}{0}
We now replace Einstein's gravitational equation
\beq {R}^0_{jk}-\tfrac{1}{2}g_{jk}R^0=\Gk T'_{jk},
\eeq{3.0}
where the $T'_{jk}$ are the elements of the stress-energy-momentum tensor $\BT'$,
and $\Gk\approx 2\times 10^{-43}s^2m^{-1}kg^{-1}$ is the gravitational constant,
with the new equation
\beq \overline{R}_{jk}-\tfrac{1}{2}g_{jk}\overline{R}=\Gk T'_{jk}.
\eeq{3.1}
This then has the equivalent form
\beq R^0_{jk}-\tfrac{1}{2}g_{jk}R^0-g_{jk}g_{mn}N^mN^n-2g_{jn}g_{km}N^mN^n
=\Gk T'_{jk},
\eeq{3.5}
or 
\beq R^0_{jk}-\tfrac{1}{2}g_{jk}R^0=\Gk T_{jk},
\eeq{3.6}
with 
\beq T_{jk}=T'_{jk}+[g_{jk}g_{mn}N^mN^n+2g_{jn}g_{km}N^mN^n]/\Gk.
\eeq{3.7}
Thus $\BT$ is the equivalent stress-energy-momentum  tensor if we were to reinterpret our equations in the format
of Einstein's original gravitational equation \eq{3.0}.
From here onwards until the last section we will assume that $\BT'=0$, i.e., that no ordinary matter is present in the region of space-time
being studied. By multiplying
\eq{3.1} by $g^{kj}$ and summing over indices we see that $\overline{R}=0$ and hence
\eq{3.5} can be rewritten as
\beq  \overline{R}_{jk}=R^0_{jk}+2g_{jk}g_{mn}N^mN^n-2g_{jn}g_{km}N^mN^n=0,
\eeq{3.7b}
or, raising indices,
\beq  \overline{R}^{jk}=\{R^0\}^{jk}+2g^{jk}g_{mn}N^mN^n-2N^jN^k=0.
\eeq{3.7c}
These equations are consistent, for example, with those of Sengupta \cite{Sengupta:2020:GTD} (see his equation (27)) which, however,
are not the same as they include an extra dimension and incorporate additional fields. 

The well known Bianchi identities between
the components of the contracted curvature tensor imply
\beq  [\{R^0\}^{jk}-\tfrac{1}{2}g^{jk} R^0]_{,k}=0,
\eeq{3.7D}
and as is well known this implies $T^{ij}{,j}=0$, reflecting conservation
of energy and momentum. Together with \eq{3.7c} and \eq{1.32} we obtain
\beq [g^{jk}g_{mn}N^mN^n+2N^jN^k]_{,k} =0 \eeq{3.7e}

We can view these
as the extra four equations needed to determine the four components of
$\BN$ in empty space.
One slightly unsatisfactory feature of the equations is that $\BN$ is only determined up a sign change. In other words,
given a solution in a spacetime region, another solution can be obtained by reversing the sign of $\BN$ within a
subregion. Thus we do not consider our theory to be complete. At the quantum Planck length scale it likely
needs modification, and the modified theory could prevent abrupt changes in the sign of  $\BN$. Alternatively, one could take the
view that there is no torsion but rather $\BN(\Bx)$ is just a vector field pervading all space. Then the sign of $\BN(\Bx)$
is immaterial, but still one would expect modifications at Planck length scale to provide a lower limit to the length scales of
``turbulence'' in the vector field  $\BN(\Bx)$ . 


\section{The weak field approximation}
\setcounter{equation}{0}

Now consider the weak field approximation where $g_{\Ga\Gb}=g^0_{\Ga\Gb}+\Gk h_{\Ga\Gb}$,
and $N^i=\sqrt{\Gk} n_i$ where $\Gk$ is a small parameter, and the $g^0_{a\Gb}$
correspond to the Minkowski metric:
\beq g_{aa}^0=\{g^0\}^{aa}=1, \quad g_{ab}^0=\{g^0\}^{ab}=0, \quad g_{a4}^0=\{g^0\}^{a4}=0, \quad g_{44}^0=\{g^0\}^{44}=-1 \eeq{3.8}
in which $a,b$ are indices taking the values $1$, $2$ or $3$ with $a\ne b$. There is some freedom in the choice of
the $h_{\Ga\Gb}$ due to the coordinate shifts that we can make to first order in $\Gk$. This freedom can be
eliminated by imposing the harmonic gauge that
\beq h^{jk}_{,k}=\tfrac{1}{2}\{g^0\}^{jk} h_{,k}, \eeq{3.8a}
in which $h=\{g^0\}^{st}h_{st}$, and $h^{jk}=\{g^0\}^{js}\{g^0\}^{kt}h_{st}$. To first order in $\Gk$ \eq{3.7c} implies
\beq 0=\overline{R}^{jk}/\Gk=-\frac{1}{2}g_{mn}^0\frac{\Md h^{jk}}{\Md x_m\Md x_n}
+2\{g^0\}^{jk}g_{mn}^0n^m n^n-2n^j n^k.
\eeq{3.8b}
Also, to first order in $\Gk$, \eq{3.7e} implies
\beq [\{g^0\}^{jk}g^0_{mn}n^mn^n+2n^jn^k]_{,k} =0. \eeq{3.8c}
Not all the 10 equations in \eq{3.8b} are independent, as a consequence of the Bianchi identities \eq{3.7D}.
To see this directly, multiply \eq{3.8b} by $g^{0}_{hj}$ and contract indices to give
\beq 0=\overline{R}/\Gk=-\frac{1}{2}g_{mn}^0\frac{\Md h}{\Md x_m\Md x_n}+6 g_{mn}^0n^m n^n,
\eeq{3.8ca}
which is also implied by taking the first order approximation to \eq{1.32}. Thus we have
\beq 0=(\overline{R}^{jk}-\tfrac{1}{2}g^{jk}\overline{R})/\Gk=-\frac{\Md}{\Md x_m\Md x_n}(h^{jk}-\tfrac{1}{2}\{g^0\}^{jk} h)
-[\{g^0\}^{jk}g_{mn}^0n^m n^n+2n^j n^k].
\eeq{3.8d}
With \eq{3.8a} we recover \eq{3.8c}. In summary, we should first use the four equations \eq{3.8c} to determine the $n^i(\Bx)$, $i=1,2,3,4$.
Then we should use the 16 equations\eq{3.8a} and \eq{3.8b}, of which only 10 are independent, to determine the 10 functions $h_{ij}(\Bx)$.
Writing out the equations \eq{3.8b} explicitly we get
\beqa \nabla^2 h^{ab}-\frac{\Md^2}{\Md t^2}h^{ab} & = & 4[\Gd_{ab}(n^2-n_4^2)-n_an_b], \nonum
\nabla^2 h^{a4}-\frac{\Md^2}{\Md t^2}h^{a4} & = & 4n_an_4, \nonum
\nabla^2 h^{44}-\frac{\Md^2}{\Md t^2}h^{ab} & = & -4n^2,
\eeqa{3.8e}
where the indices $a$ and $b$ take values from $1$ to $3$, $n^2=n_1^2+n_2^2+n_3^2$ and $n_i=g_{ij}^0n^j$. As we have used the harmonic gauge there is the additional
restriction that the $h^{jk}$ satisfy \eq{3.8a}, i.e. that
\beqa h^{a1}_{,1}+ h^{a2}_{,2}+ h^{a3}_{,3}+ h^{a4}_{,4}& = & \tfrac{1}{2}(h^{11}+ h^{22}+ h^{33}- h^{44})_{,a},\quad a=1,2,3,  \nonum
 h^{41}_{,1}+ h^{42}_{,2}+ h^{43}_{,3}+ h^{44}_{,4} & = & -\tfrac{1}{2}(h^{11}+ h^{22}+ h^{33}- h^{44})_{,4}.
\eeqa{3.8f}

The identities \eq{3.8c} imply $T^{ij}_{,j}=0$ with, to zeroth order in $\Gk$,
\beqa T^{aa}& = & 2n_a^2+n^2-n_4^2,\quad T^{ab}=2n_an_b,
\nonum
T^{44}& = & 3n_4^2-n^2,\quad T^{a4}=-2n_an_4.
\eeqa{3.9}
Equivalently, the matrix $\BT$ with elements $T^{ij}$ takes the block form:
\beq \BT = \bpm 2\Bn\otimes\Bn +(n^2-n_4^2)\BI & -2n_4\Bn \cr -2n_4\Bn^T & 3n_4^2-n^2\epm, 
\eeq{3.9a}
where $\Bn^T$ is the row vector that is the transpose of $\Bn$, defined as $\Bn=(n_1,n_2,n_3)$.

\section{Subluminal, Luminal and Superluminal Regions
of Spacetime}
\setcounter{equation}{0}
In this section we do not make the weak field approximation, but we consider any point $P$ in spacetime and
choose the  Minkowski metric \eq{3.8} at that point.
\subsection{Subluminal Regions and the equivalent perfect fluid
  with negative energy that occupies them}
Consider a region where $k=n_4^2-n^2>0$. We call such a region
a subluminal region. Define the 4-velocity $\BV$
with components
\beq V_a=n_a/\sqrt{k},\quad V_4=n_4/\sqrt{k} \eeq{3.11}
satisfying $V_1^2+V_2^2+V_3^2-V_4^2=-1$. In terms of this velocity \eq{3.9} implies
\beqa T^{aa}& = & (2V_a^2-1)k, \quad  T^{ab}=2V_aV_bk ,\nonum
T^{44}& = & (2V_4^2+1)k,\quad T^{a4}=2V_aV_4k.
\eeqa{3.12}
By comparison, a perfect fluid moving with 4-velocity $\BV$ has
\beqa T^{aa}& = & (\Gm_{0}+p)V_a^2+p,\quad
T^{ab}=(\Gm_{0}+p)V_aV_b ,\nonum
T^{44}& = &(\Gm_{0}+p)V_4^2-p,\quad
T^{a4} = (\Gm_{0}+p)V_aV_4,
\eeqa{3.13}
where $p=p$ is the pressure and $\Gm_0$ is the rest density (in the frame
with the same velocity as the fluid). Thus $T$ corresponds to a fluid
with
\beq p=-\Gm_0/3,\quad \Gm_0=3k. \eeq{3.14}
Note that in this case it always possible to choose a moving frame of
reference with respect to which the fluid is not locally moving, i.e.
$n^2=0$.
\subsection{Superluminal Regions and the equivalent substance
  with negative mass that occupies them}
Consider those regions where $k=n_4^2-n^2<0$, which we call superluminal.
Then it is impossible to move to a reference frame such that $n^2=0$ at a given point. Rather we can move to a frame where
$n_4=0$ at this point. In this frame
\beqa T^{aa}& = & 2n_a^2+n^2,\quad T^{ab}=T_{ab}+2n_an_b,
\nonum
T^{44}& = & -n^2,\quad T^{a4}=0.
\eeqa{3.15}
This corresponds to some sort of substance that, in this frame, has no momentum, a negative mass density $-n^2$ and a stress
\beq \BGs=-n^2\BI-2\Bn\otimes\Bn, \eeq{3.16}
corresponding to a pressure of $n^2$ and an additional uniaxial compression
in the direction $\Bn$.
\subsection{Luminal Regions}

Finally, consider the regions where $k=n_4^2-n^2=0$, which we call luminal.
Then
\beq T^{aa} = 2n_a^2,\quad T^{ab}=2n_an_b,\quad
T^{44} = 2n_4^2,\quad T^{a4}=-2n_an_4.
\eeq{3.18}
Clearly a luminal boundary or luminal region must separate regions that
are subluminal or superluminal. In a luminal region one cannot move to a frame
where $n^2=0$, nor where $n_4^2=0$, unless both are zero. The momentum density,
mass density, and stress are non-zero everywhere, except where the torsion field vanishes. 

\section{Some solutions for the torsion field in the weak field approximation}
\setcounter{equation}{0}
Let us consider solutions of $T^{ij}_{,j}=0$ in a flat metric given by \eq{3.8}. Using \eq{3.9a} we obtain
\beqa 0 & = & \frac{\Md}{\Md t}[3n_4^2-n^2]
-2\Div(n_4\Bn) ,\nonum
0 & = &  \Div(\Bn\otimes\Bn)-\frac{\Md(n_4\Bn)}{\Md t}+\tfrac{1}{2}\Grad(n^2-n_4^2) \nonum
  & = & (\Bn\cdot\Grad)\Bn +\Bn\Div\Bn-\frac{\Md(n_4\Bn)}{\Md t}+\tfrac{1}{2}\Grad(n^2-n_4^2),
\eeqa{4.b}
where the first equation represents conservation of energy and the second
balance of forces.

In the superluminal regions if we look for solutions where
$n_4=0$ globally and not just at one point, then conservation of energy implies
that $n^2$ must not vary with time, and balance of forces implies
\beq \Grad(n^2)+2\Div(\Bn\otimes\Bn)=0.
\eeq{3.17}
This provides 3 equations to be satisfied by the three functions $n_a(x_1,x_2,x_3,t)$,
$a=1,2,3$. There is a manifold of functions satisfying \eq{3.17}, and we
can choose any trajectory $\Bn(x_1,x_2,x_3,t)$ that lies on this
manifold and is such that $n^2(x_1,x_2,x_3)=\Bn(x_1,x_2,x_3,t)\cdot\Bn(x_1,x_2,x_3,t)$ is independent of time.
It seems likely that this second condition will generally force $\Bn(x_1,x_2,x_3,t)$ to be independent of time.

In luminal regions where $n^2-n_4^2=0$ we can use this identity to
 eliminate $n_4$ from \eq{4.b} and get
\beqa 0 & = & \frac{\Md n^2}{\Md t} \pm \Div(|\Bn|\Bn) ,\nonum
0 & = &  \Div(\Bn\otimes\Bn) \pm \frac{\Md(|\Bn|\Bn)}{\Md t} ,\nonum
& = & (\Bn\cdot\Grad)\Bn +\Bn\Div\Bn \pm \frac{\Md(|\Bn|\Bn)}{\Md t},
\eeqa{4.c}
where the plus or minus sign is taken according to whether $n_4=\pm |\Bn|$.
In the special case where $n_2=n_3=0$ (after making a spatial rotation
if necessary) we get $n_4=n_1$ (or $n_4=-n_1$) and \eq{4.c} reduces to the single equation
\beq \frac{\Md n_1}{\Md t}=\frac{\Md n_1}{\Md x_1} \eeq{3.20}
to be satisfied by the function $n_1(x_1,x_2,x_3,t)$,
describing a wave propagating at the speed of light in the direction of the $x_1$-axis. We call them localized longitudinal torsion waves,
longitudinal because $\Bn$ is aligned with the direction of propagation.

\subsection{Plane Wave Solutions}
Here we consider plane wave solutions to the equations in the weak field approximation.
It is to be emphasized that since the equations are non-linear, specifically quadratic
in $\Bn$, one cannot generally superimpose our plane wave solutions to get another solution.

The simplest case is when the fields only depend on say $x_1$. Then we
we deduce that $T^{1j}$ is a constant, i.e.
\beq 3n_1^2+n_2^2+n_3^2-n_4^2=k_1,\quad
n_1n_2=k_2,\quad n_1n_3=k_3,\quad n_1n_4=k_4,
\eeq{4.1}
where the $k_i$ are constants. Multiplying the first equation by $n_1^2$ we obtain
\beq n_1^4= (k_1-k_2^2-k_3^2+k_4^2)/3, \eeq{4.1a}
which requires the constants  $k_i$ to be such that right hand side is non-negative.
Thus $n_1^2$ is constant, and the last three equations in \eq{4.1} imply
that $n_2^2$, $n_3^2$, and $n_4^2$ are constants too, unless $n_1^2=0$. So the
only interesting case is when $n_1^2=0$, implying that $k_2=k_3=k_4=0$.
Additionally, \eq{4.1a} implies that $k_1=0$ too. The first equation in
\eq{4.1} forces us to be in the luminal region where $n^2-n_4^2=0$. Thus $n_2(x_1)$ and $n_3(x_1)$
can be chosen arbitrarily and determine $n_4^2=n^2$. In particular, one may choose $n_2(x_1)$ and $n_3(x_1)$
to be zero outside an interval of values of $x_1$.
In a frame of reference
moving with velocity $-v_1$ in direction $x_1$ this will look like a wave pulse
traveling a velocity $v_1$ as all the field components will be functions
of $x_1-v_1t$. We call them localized transverse torsion waves,
transverse because $\Bn$ is perpendicular to the wave front. Unlike longitudinal torsion waves, which
can only travel at the speed of light, these can have any velocity less than $c$. 

Similarly, when the fields only depend on
$t=x_4$ we deduce that $T^{4j}$ is a constant, i.e.
\beq n_4\Bn=\Bk',\quad 3n_4^2-n^2=k_4',
\eeq{4.2}
in which $\Bn=(n_1,n_2,n_3)$ and $n^2=\Bn\cdot\Bn$ and
where $k_4'$ and $\Bk'=(k'_1,k'_2,k'_3)$ are constants. Multiplying the last formula
by $n_4^2$ shows that
\beq n_4^4=(k_4'-\Bk'\cdot\Bk')/3 \eeq{4.3}
is constant, implying that $n_1^2$, $n_2^2$, and $n_3^2$ are constant too unless $n_4^2=0$. When
$n_4^2=0$ then $\Bk'=0$ and \eq{4.3} implies $k_4'=0$. The last formula in \eq{4.2}
then forces $\Bn=0$. So there are no non-trivial solutions when the torsion field only depends
on $t$.

\subsection{Solutions with cylindrical symmetry, including torsion-rolls}
Consider cylindrical coordinates $(r,\Gt,z,t)$ taking $r$ to be the radial
distance from the $z$-axis,
$\Gt$ to be the angular variable, and $t$ to be the time. We seek solutions where $\Bn=(n_r,n_\Gt,n_z)$ and $n_4$
only depend on $r$, so that
\beqa \Div(n_4\Bn) & = & \frac{1}{r}\frac{d(rn_4n_r)}{d r}\hat\Br, \nonum
(\Bn\cdot\Grad)\Bn & = & \left(n_r\frac{d n_r}{d r}-\frac{n_r^2}{r}\right)\hat\Br
+\left(n_r\frac{d n_\Gt}{d r}+\frac{n_rn_\Gt}{r}\right)\hat\Gt
+\left(n_r\frac{d n_z}{d r}\right)\hat\Bz, \nonum
\Bn(\Div\Bn)& = & \frac{1}{r}\frac{d (r n_r)}{d r}(n_r\hat\Br+n_\Gt\hat\Gt+n_z\hat\Bz), \nonum
\tfrac{1}{2}\Grad(n^2-n_4^2) & = & \tfrac{1}{2}\left[\frac{d}{d r}(n^2-n_4^2)\right]\hat\Br,
\eeqa{3-1}
 where we have used the standard formulas for the gradient, divergence, and $\Bn\cdot\Grad$ in cylindrical coordinates.
Then the conservation laws \eq{4.b} take the form
\beqa 0 & = & \frac{1}{r}\frac{d (rn_rn_4)}{d r}, \nonum
  0 & = & \frac{n_r^2-n_\Gt^2}{r}
  +\tfrac{1}{2}\frac{d}{d r}\left[2n_r^2+n^2-n_4^2\right], \nonum
  0 & = & \frac{2n_r n_\Gt}{r}+\frac{d (n_r n_\Gt)}{d r},\nonum
  0 & = & \frac{n_r n_z}{r}+\frac{d (n_rn_z)}{d r}.
  \eeqa{3.30}
If we consider an interface at a constant radius $r=r_0$, with outwards unit normal $\hat\Br$, then the weak form
of the equations $T^{ij}_{,j}=0$ imply the jump conditions on the elements $T^{ij}$ that
  $$ \BT\bpm \hat\Br \\ 0 \epm $$
  must be continuous across the interface, where $\BT$ is given by \eq{3.9a}. 
  This implies that the quantities
  \beq k_4=n_4n_r,\quad k_\Gt=n_\Gt n_r, \quad k_z=n_zn_r, \quad k_r=3n_r^2+n_\Gt^2+n_z^2-n_4^2
  \eeq{3.30a}
  must all be continuous across the interface $r=r_0$. Multiplying the last equation by $n_r^2$ we see
  that
  \beq n_r^4=(k_r-k_\Gt^2-k_z^2+k_4^2)/3 \eeq{3.30b}
  must be continuous too, and the first three equations imply that all components of $(\Bn,n_4)$ are continuous across the
  interface, up to a change of sign, unless $n_r^2=0$ at the interface. If $n_r^2$ is zero at the interface it follows that
  $k_4= k_\Gt=k_z=0$ at the interface, and \eq{3.30b}
  implies that additionally $k_r=0$. So, across $r=r_0$, any jumps in $n_\Gt(r,t)$, $n_z(r,t)$ and $n_4(r,t)$ that  maintain
  the continuity of $n^2-n_4^2$ are possible provided $n_r(r,t)$ is continuous and $n_r(r_0,t)=0$. 

  The first, third, and last equations in \eq{3.30} imply
  \beq rn_rn_4=k_4,\quad rn_rn_z=k_z,\quad r^2n_rn_\Gt=k_\Gt
  \eeq{3.30c}
  where $k_4$, $k_z$, and $k_\Gt$ are constants. In the case $n_r=0$, all are satisfied with $k_4=k_z=k_\Gt=0$. The remaining 
second equation in \eq{3.30} becomes
  \beq  \frac{d}{d r}\left[n^2-n_4^2\right]=\frac{2n_\Gt^2}{r}.
  \eeq{3.31}
  Thus there is only one constraint among the three functions $n_\Gt(r)$, $n_z(r)$, and $n_4(r)$. We see that
  $n^2-n_4^2$ must monotonically increase with $r$, in a manner controlled by $n_\Gt^2(r)$
  and if it tends to zero at infinity, then $n^2-n_4^2$
  must be negative for all $r$, corresponding to a subluminal region. If $\Bn$ and $n_4$ vanish outside a
  certain radius then we call this solution a torsion-roll. Physically, the pressure increases to larger negative
  values as the radius decreases and its gradient provides the centripetal force that holds the ``fluid''
  circulating around the $z$-axis with a velocity governed by $n_\Gt$. In a moving frame of reference, which is not moving in
  the $z$-direction, the torsion-roll will appear to be moving.

  Of course, if $n^2-n_4^2$ is constant and positive
  outside a certain radius (corresponding, for example, to a superluminal region where say $n_z$ is constant and
  $n_\Gt=n_4=0$) then $n^2-n_4^2$ can remain positive for all r, or can transition from positive to negative
  values at a particular radius. This example demonstrates that transitions between subluminal and superluminal
  regions are possible.

  Alternatively, if $n_r$ is non-zero, then \eq{3.30c} implies
  \beq n_4=k_4/(rn_r),\quad n_z=k_z/(rn_r),\quad n_\Gt=k_\Gt/(r^2n_r).
  \eeq{3.1a}
  Substituting these in the second equation in \eq{3.30} yields
  \beq \frac{d s}{d r}=\frac{2s(3k_\Gt^2-r^4s^2-2kr^2)}{r(3r^4s^2-k_\Gt^2+kr^2)},\quad\text{where}\quad s=n_r^2,\quad k=k_4^2-k_z^2.
  \eeq{3.37}
  This gives us a flow-field in the $(r,s)$ phase plane.
   Note that \eq{3.37} remains invariant under the transformation
  \beq r\to \Gl_1 r,\quad s \to \Gl_2 s, \quad k_\Gt^2\to \Gl_1^4\Gl_2^2 k_\Gt^2, \quad k\to \Gl_1^2\Gl_2^2 k. \eeq{3.39}
  Thus, without loss of generality, we may by rescaling any solution take $k_{\Gt}$ to be 0 or 1 and $k$ to
  be $-1$, 0, or 1. If $k=0$  then there is essentially just one solution:
  $s_0(r)$ satisfying $s_0(1)=1$ with all other solutions (with $k_\Gt=1$) taking the form $s(r)=\Gl^2s_0(\Gl r)$,
  parametrized by $\Gl$. The solutions for $s_0(r)=n_r^2(r)$ and $n_\Gt^2(r)=1/(r^4n_r^2)$ are shown in \fig{k=0} along with the flow field. One can
  see that the solution does not exist below a critical value of $r$, which looks unsatisfactory. This critical radius is associated with
  the vanishing of the denominator in \eq{3.37}. 
  \begin{figure}[ht!]
	\centering
	\begin{subfigure}{.45\linewidth}
		\includegraphics[width=\textwidth]{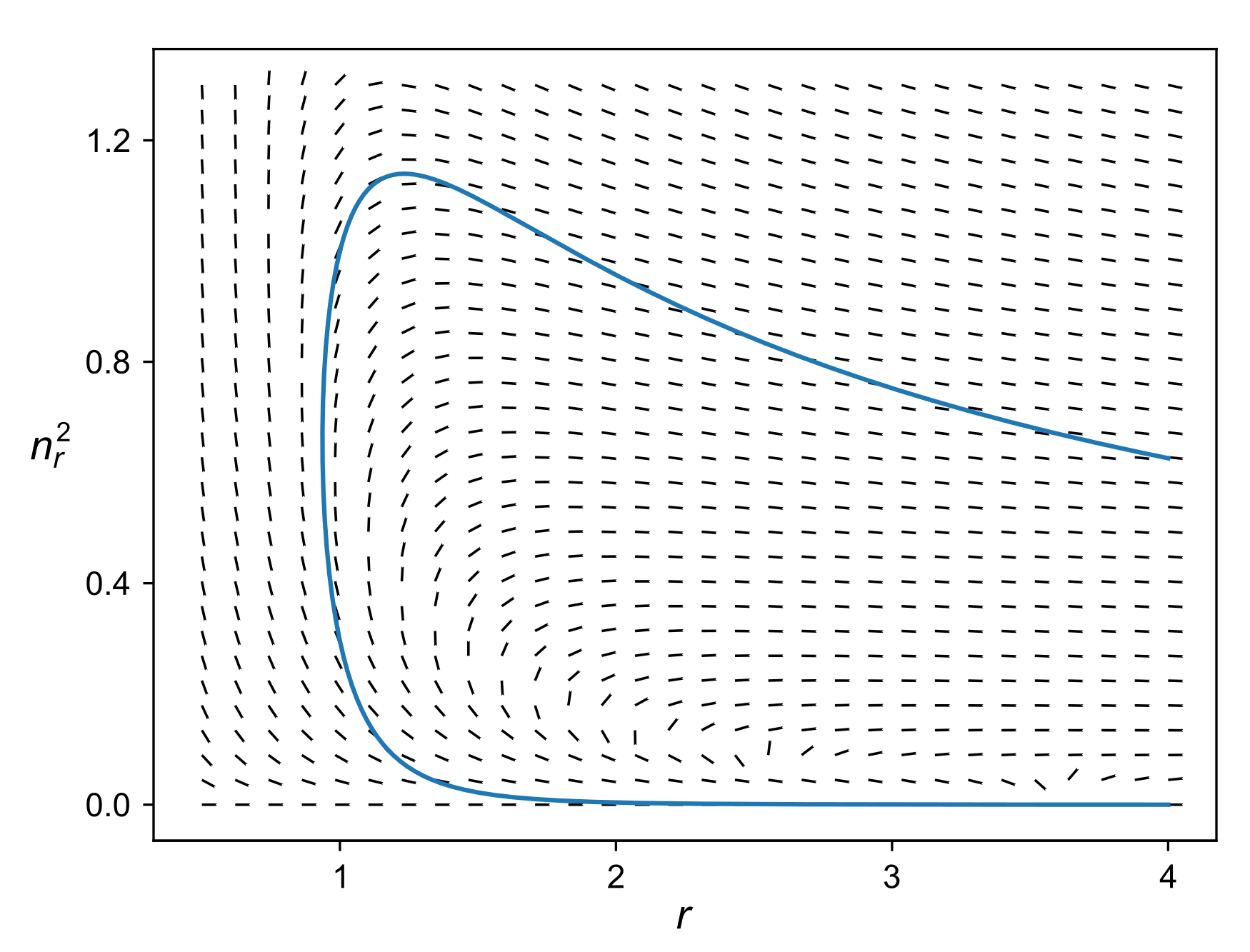}
		\caption{The flow field when $k=0$ and $k_\Gt=1$, and the particular solution satisfying $n_r^2=1$ when $r=1$}
	\end{subfigure}
	\hskip2em
	\begin{subfigure}{.45\linewidth}
		\includegraphics[width=\textwidth]{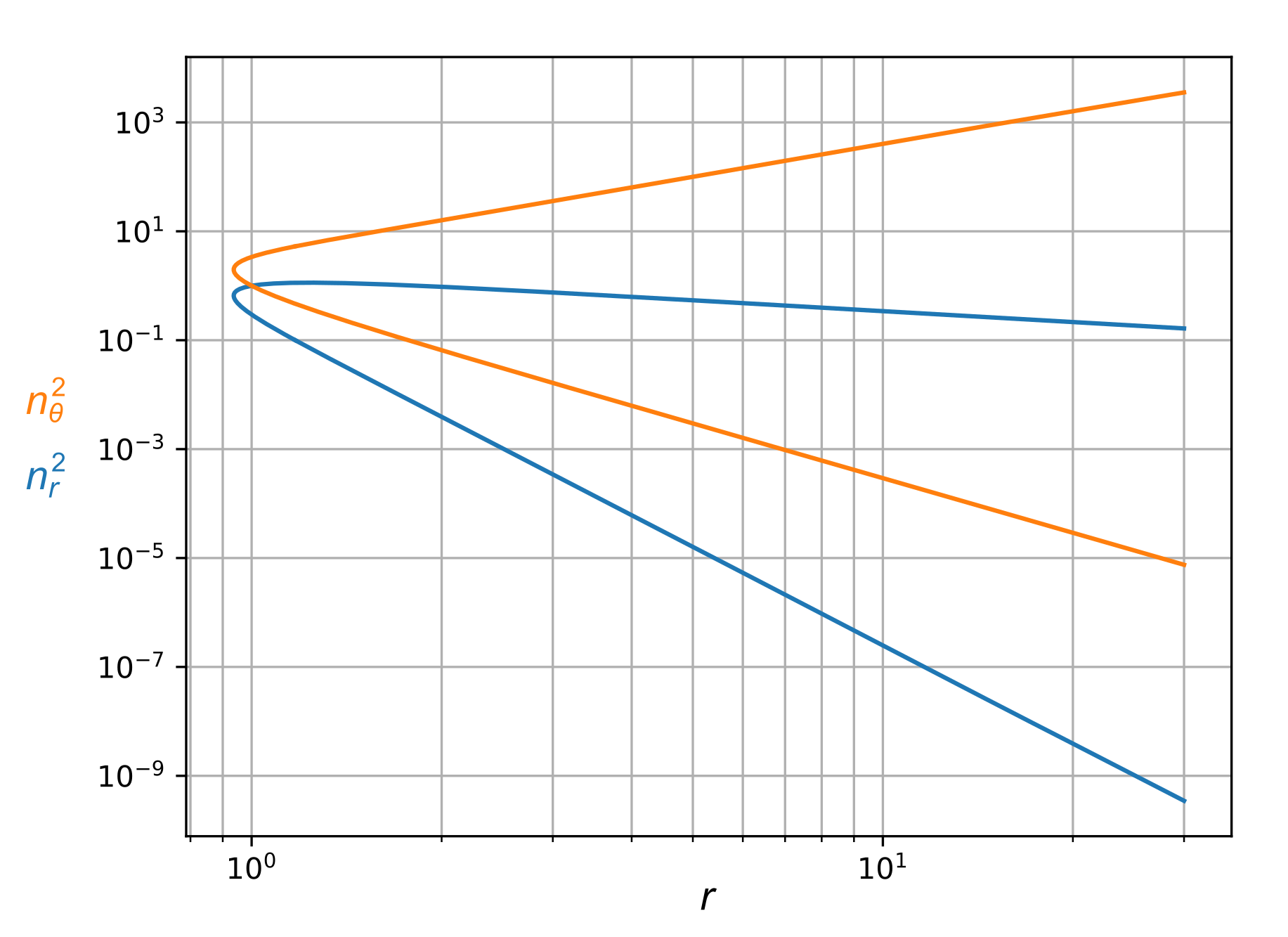}
		\caption{The same solution for $n_r^2$ on a log-log plot and the accompanying function $n_\Gt^2=1/(r^4n_r^2)$.}
	\end{subfigure}
	\caption{Solution for the torsion field with cylindrical symmetry with $n_r\ne 0$, $k=0$ and $k_\Gt=1$}
	\labfig{k=0}
      \end{figure}

      To obtain satisfactory solutions that exist for all $r\ne 0$ one may take $k_\Gt=0$ and $k=1$ to avoid the denominator in \eq{3.37} vanishing except at
      $r=0$. Then \eq{3.37} reduces to
\beq \frac{d s}{d r}=-\frac{2s(r^2s^2+2)}{r(3r^2s^2+k)},\quad\text{where}\quad s=n_r^2.
\eeq{3.37a}
 There is again essentially just one solution:
  $s_0(r)$ satisfying $s_0(1)=1$ with all other solutions (with $k=1$) taking the form $s(r)=\Gl s_0(\Gl r)$,
  parametrized by $\Gl$. The solution is graphed in \fig{k=1}. There is a singularity at $r=0$ and while $n_r^2(r)$ goes rapidly to zero
  as $r\to\infty$, $n_4^2(r)$ and $n_z^2(r)$ (unless it is zero) diverge to $\infty$ as $r\to\infty$. This solution is satisfactory once one
  takes into account that the weak field approximation is not valid near the singularity at $r=0$, nor as $r\to\infty$, and one should
  use the full equations \eq{3.7c} there. For this example with $k=1$ and $k_\Gt=0$,
  it is interesting that there is a transition from a superluminal region inside to a subluminal region outside according to the
  sign of
  \beq n^2-n_4^2=n_r^2+\frac{k_z^2}{r^2n_r^2}-\frac{k_4^2}{r^2n_r^2}=n_r^2-\frac{1}{r^2n_r^2},
  \eeq{3.37b}
  which is also plotted in \fig{k=1}.
  
  \begin{figure}[ht!]
	\centering
	\begin{subfigure}{.45\linewidth}
		\includegraphics[width=\textwidth]{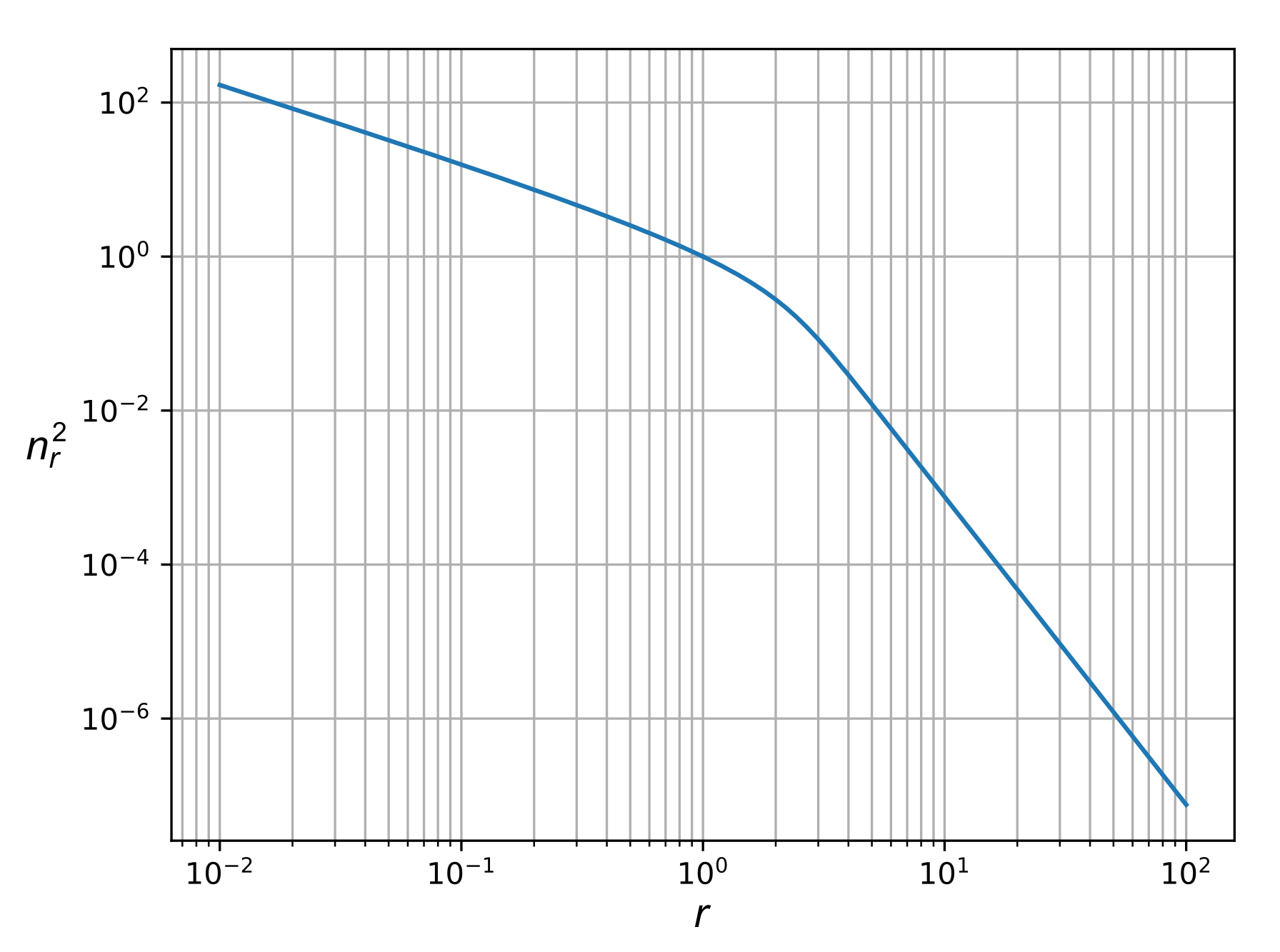}
		\caption{The graph of $n_r^2=1$ showing its divergence as $r\to 0$}
	\end{subfigure}
	\hskip2em
	\begin{subfigure}{.45\linewidth}
		\includegraphics[width=\textwidth]{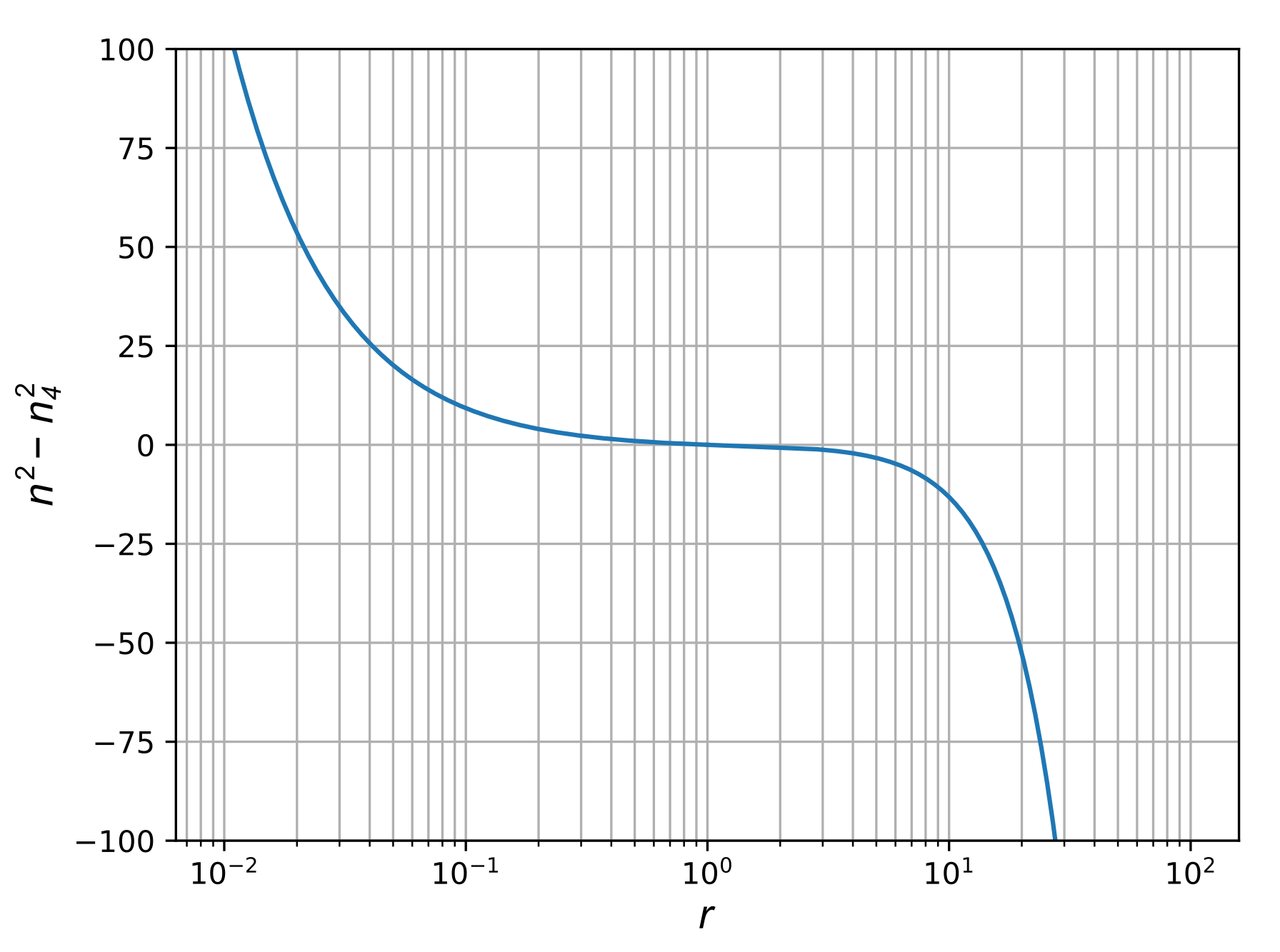}
		\caption{The plot of $n^2-n_4^2=n_r^2-1/(r^4n_r^2)$ showing a transition from superluminal to subluminal as $r$ increases}
	\end{subfigure}
	\caption{Solution for the torsion field with cylindrical symmetry with $n_r\ne 0$, $k=1$ and $k_\Gt=0$}
	\labfig{k=1}
      \end{figure}

\section{Extension of the Schwarzschild solutions with
spherical symmetry}
\setcounter{equation}{0}
Here we generalize Schwarzschild's solution for a spherically symmetric
metric solving Einstein's equations in the absence of matter. The important point is that in appropriate limits some of the solutions here approach
the Schwarzschild solution. Consequently, existing experimental results of black holes do not invalidate
our theory, but rather place constraints on the magnitude of the torsion
field. This magnitude should be tied to the radius of the universe, and hence
to the critical acceleration in MOND. Thus experiments in the near
vicinity of a star or black hole would not typically reveal the difference with
Schwarzschild's solution. We have not explored the situation regarding rotating black holes.

As shown by Schwarzschild the metric in ``polar'' coordinates spherically
symmetric  about the origin must be of the form
\beq ds^2=a\,dr^2+r^2(d\Gt^2+\sin^2\Gt\,d\Gf^2)-b\,dt^2, \eeq{5.1}
in which $a$ and $b$ are functions of $r$ and $t$. Here we
look for solutions where they are functions of $r$ only.
Setting $x_1=r$, $x_2=\Gt$, $x_3=\Gt$, $x_4=t$ allows us to use
\eq{5.1} to identify the coefficients
\beq g_{11}=a,\quad g_{22}=r^2,\quad g_{33}=r^2\sin^2\Gt,\quad g_{44}=-b.
\eeq{5.2}
From \eq{3.7b} we obtain the ten equations
\beqa
0 & = &\overline{R}_{11}=\frac{a'}{ar}+\frac{a'b'}{4ab}+\frac{(b')^2}{4b^2}-\frac{b''}{2b}+2a[r^2(N^2)^2+r^2\sin^2\Gt(N^3)^2-b(N^4)^2],\nonum
0 & = &\overline{R}_{22}=1-\frac{1}{a}+\frac{ra'}{2a^2}-\frac{rb'}{2ab}+2r^2[a(N^1)^2+r^2\sin^2\Gt(N^3)^2-b(N^4)^2],\nonum
0 & = &\overline{R}_{33}=[1-\frac{1}{a}+\frac{ra'}{2a^2}-\frac{rb'}{2ab}]\sin^2\Gt+2r^2\sin^2\Gt[a(N^1)^2+r^2(N^2)^2-b(N^4)],\nonum
0 & = &\overline{R}_{44}=\left(\frac{b'}{ar}+\frac{b''}{2a}-\frac{(b')^2}{4ab}-\frac{a'b'}{4a^2}\right)-2b[a(N^1)^2+r^2(N^2)^2+r^2\sin^2\Gt(N^3)^2]  ,\nonum
0 & = &\overline{R}_{mn}=-2g_{mm}g_{nn}N^mN^n\quad \text{for all }m,n\quad \text{with  }m\ne n,\quad \text{no sum on } m,n,
\eeqa{5.3}
where the terms not involving $\BN$ can be identified with the standard formulas for the elements $R^0_{ij}$ that are zero when $i\ne j$. Here
 differentiation with respect to $x_1=r$ is denoted by the prime, with
 the double prime denoting the second derivative. The second and third equations and the last equation force $N^2=N^3=0$ which is not
 surprising considering the symmetry of the problem. Two possibilities remain: either $N^1=0$ or $N^4=0$. The first case
corresponds to a subluminal solution and the second to a superluminal solution.

Let us consider first the case where $N^1=N^2=N^3=0$. Multiplying the second last equation in \eq{5.3} by $a/b$ and adding it to the first gives
\beq \frac{a'}{a}+\frac{b'}{b}-2q=0\quad \text{where  }q=rab(N^4)^2\geq 0. \eeq{5.4}
The second equation in \eq{5.3} implies
  \beq \frac{a'}{a}-\frac{b'}{b}+2(a-1)/r-4q=0. \eeq{5.5}
Adding and subtracting these equations gives 
\beqa a'/a&= & \frac{1}{r}-\frac{a}{r}+3q, \nonum
b'/b& = &\frac{a}{r}-\frac{1}{r}-q.
\eeqa{5.6}
Multiplying the last by $br$, differentiating it, and using the result to eliminate $b''$ from the first equation in \eq{5.3} yields
\beq q' = 2q^2+\frac{q}{r}. \eeq{5.7}
This has the solution
\beq q=\frac{\Ga^2r}{1-\Ga^2r^2}, \eeq{5.8}
where $\Ga$ is a constant. Also, by replacing $q$ with $rab(N^4)^2$ one obtains
\beqa 2q^2+\frac{q}{r}=q'& = &ab(N^4)^2+(ra'/a)ab(N^4)^2+(rb'/b)ab(N^4)^2+rab\frac{(N^4)^2}{dr} ,\nonum
&=& \frac{q}{r}[1+(1-a+3qr)+(a-1-qr)]+q\frac{(N^4)^2}{dr}=\frac{q}{r}+2q^2+q\frac{(N^4)^2}{dr}.
\eeqa{5.8A}
This implies that $(N^4)^2$ is a constant that we call $\Gb^2$, giving
\beq \frac{a}{r}=\frac{q}{br^2(N^4)^2}=\frac{\Ga^2}{br\Gb^2(1-\Ga^2r^2)}.
\eeq{5.9A}
Substituting this back in the second equation in \eq{5.6} gives the linear first order differential equation
\beq \frac{db}{dr} +b\left[\frac{1}{r}+\frac{\Ga^2}{1-\Ga^2r^2}\right]=   \frac{\Ga^2}{\Gb^2r(1-\Ga^2r^2)}.
\eeq{5.10A}
Multiplying both sides by the integrating factor of $r/\sqrt{1-\Ga^2r^2}$ gives
\beq \frac{d}{dr}\left[br/\sqrt{1-\Ga^2r^2}\right]
=\frac{\Ga^2}{\Gb^2(1-\Ga^2r^2)\sqrt{1-\Ga^2r^2}}.
\eeq{5.11A}
Integrating both sides and recalling \eq{5.9A} we get
\beqa b & = &\frac{\Ga^2}{\Gb^2}-2m\frac{\sqrt{1-\Ga^2r^2}}{r}, \nonum
      a & = & \frac{\Ga^2}{b\Gb^2(1-\Ga^2r^2)},
\eeqa{5.12A}
where $m$ is a constant of integration. In particular, with $\Ga^2=\Gb^2$ this becomes
\beqa b & = &1-2m\frac{\sqrt{1-\Ga^2r^2}}{r}, \nonum
      a & = & \frac{1}{b(1-\Ga^2r^2)},
\eeqa{5.13A}
which in the limit $\Ga\to 0$ reduces to the familiar Schwarzschild solution
\beq a=\frac{1}{1-2m/r},\quad b=1-2m/r, \eeq{5.9}
that becomes Euclidean at large $r$. Once we allow nonzero $\Ga$, the
space is no longer Euclidean at large $r$ but it still has a black hole
at the center, with $a$ diverging when $r=2m\sqrt{1-\Ga^2r^2}$ and at $r=1/\Ga^2$, the latter
corresponding to the closed universe studied in the next section.

  Now, consider the second possibility that $N^2=N^3=N^4=0$. Again multiplying the second last equation in \eq{5.3} by $a/(b)$ and adding it to the first gives
  \beq \frac{a'}{a}+\frac{b'}{b}-2w=0\quad \text{where  }w=ra^2(N^1)^2\geq 0. \eeq{5.20}
  Also the second equation in \eq{5.3} implies
  \beq \frac{a'}{a}+\frac{b'}{b}+2(a-1)/r+4w=0. \eeq{5.21}
Adding and subtracting these equations gives 
\beqa a'/a&= & \frac{1}{r}-\frac{a}{r}-w, \nonum
b'/b& = &\frac{a}{r}-\frac{1}{r}+3w.
\eeqa{5.22}
Multiplying the last by $br$, differentiating it, and using the result to eliminate $b''$ from the first equation in \eq{5.3} yields
\beq  w' = -2w^2+w(1-\tfrac{4}{3}a)/r. \eeq{5.23}
The equations \eq{5.22} and \eq{5.23} appear to have no simple analytic solution. One may eliminate $a(r)$ from the two equations 
that do not involve $b(r)$ to obtain
\beq \frac{w''}{w}=\frac{7(w')^2}{4w^2}-\frac{3w'}{2wr}
    -\frac{2w}{r}+w^2-\frac{1}{4r^2},
    \eeq{5.23a}
and from a solution $w(r)$, \eq{5.23} easily gives $a(r)$.    
Alternatively, one may eliminate $w(r)$ from these equations to obtain
\beq \frac{v''}{v}=\frac{3(v')^2}{v^2}+\frac{v'}{vr}+(5v'+2v^2)/3+v/r.
\eeq{5.24}
where $v=a/r$, and given a solution $a(r)=rv(r)$, the first equation in \eq{5.22} yields $w(r)$. In either case
$b(r)$ is found by integrating the last equation in \eq{5.22}.  Note that if $b(r)$ is a solution then so will be
    $\Gl^2 b(r)$ for any constant $\Gl$, i.e. $b(r)$ is only determined up to a multiplicative constant. This reflects the fact
    that we are free to rescale the time coordinate, replacing $t$ by $t/\Gl$ in \eq{5.1}.

    Rather than dealing with these second order equations for $w(r)$ and $v(r)$ one can numerically solve \eq{5.22} and \eq{5.23} directly. \fig{Sch} shows some
    typical solutions, excluding unphysical examples where say $a(r)$ or $w(r)$ remain negative for all $r$

    \begin{figure}[ht!]
	\centering
	\begin{subfigure}{.45\linewidth}
		\includegraphics[width=\textwidth]{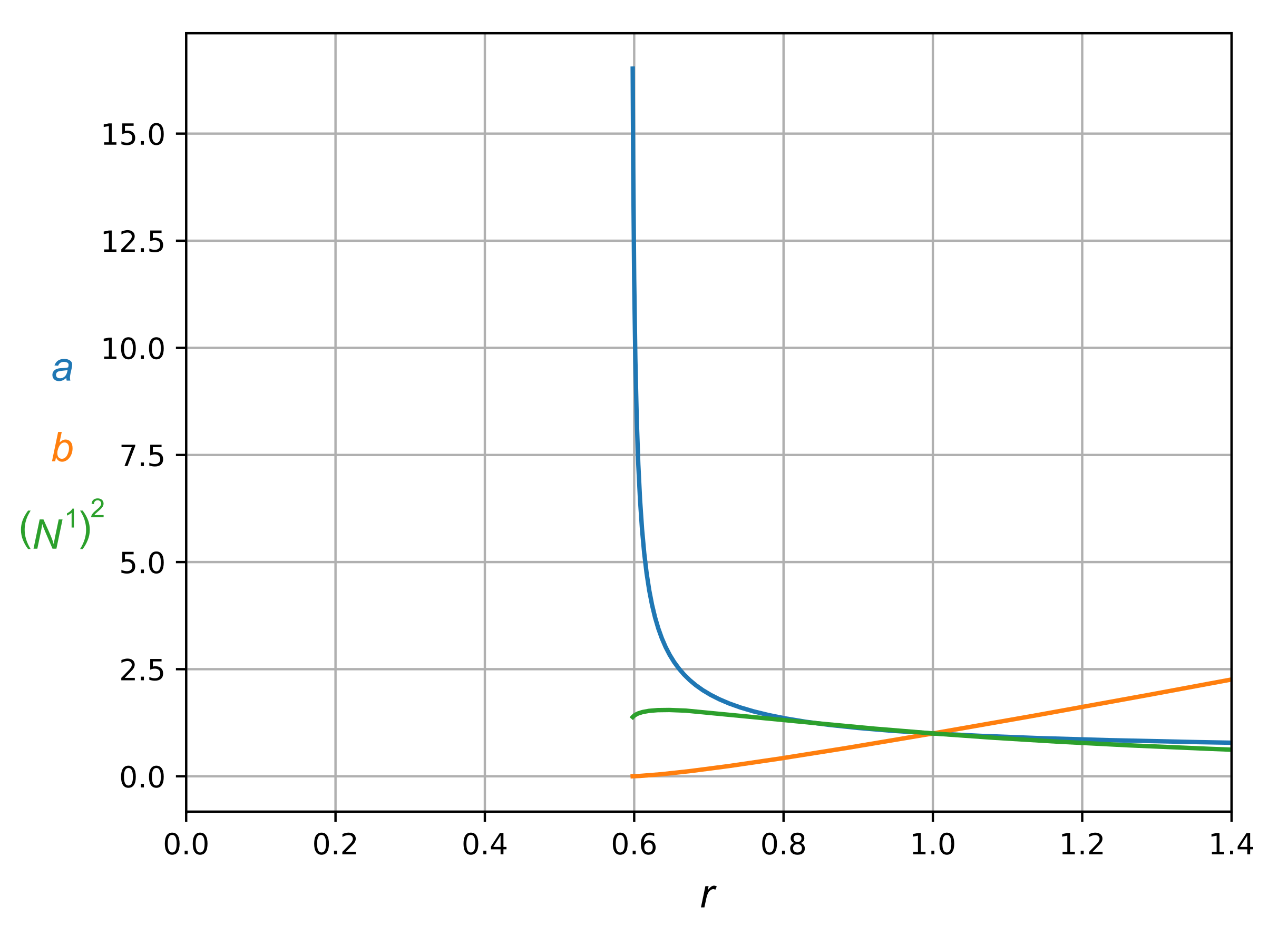}
		\caption{Graph with $w(1)=a(1)=b(1)=1$ showing a ``black-hole'' type singularity at $r=0.5959$.}
	\end{subfigure}
	\hskip2em
	\begin{subfigure}{.45\linewidth}
		\includegraphics[width=\textwidth]{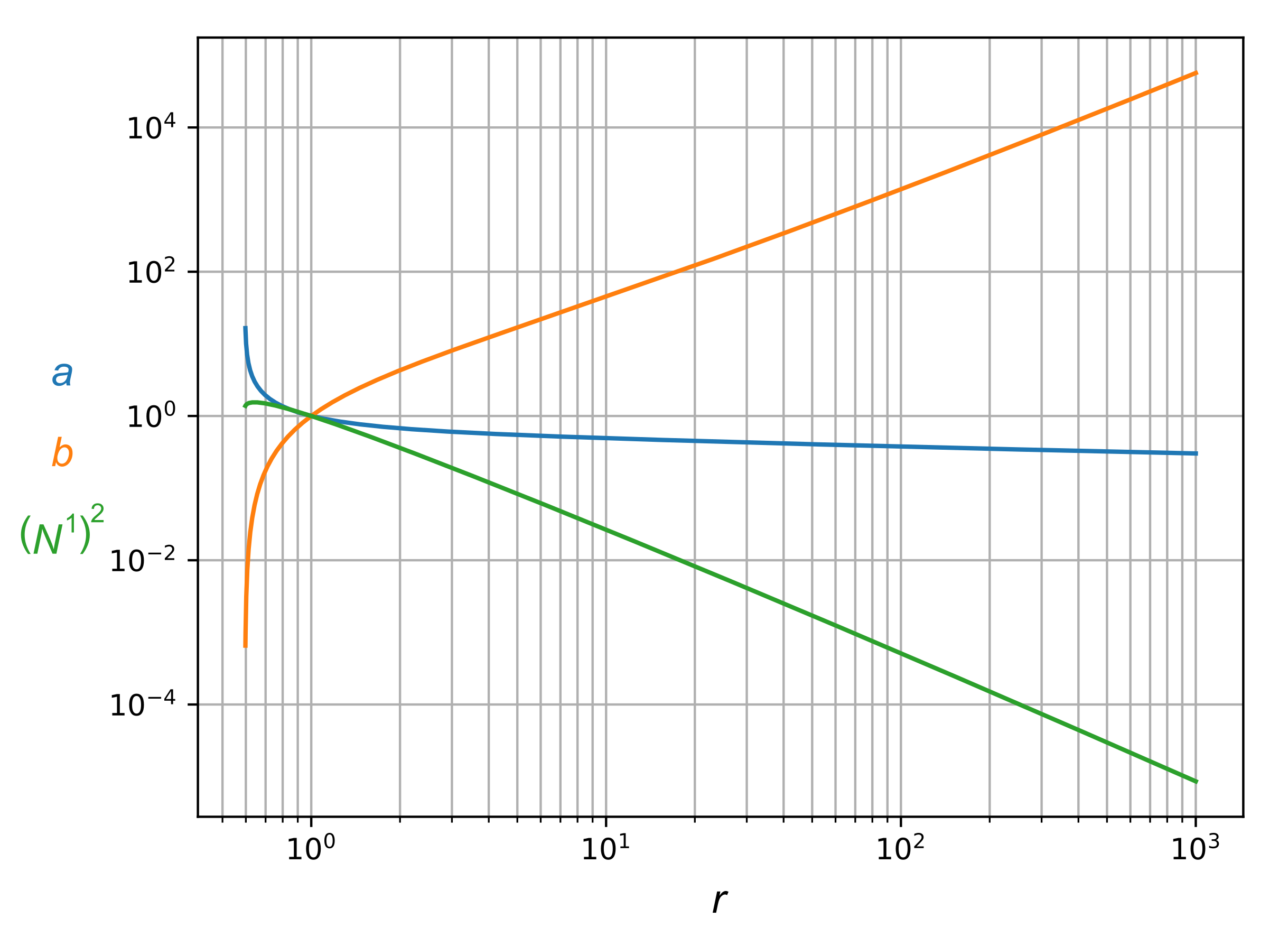}
		\caption{Same as for (a) but on a log-log plot. Note the blow up of $b(r)$ as $r\to\infty$.}
	\end{subfigure}
	\begin{subfigure}{.45\linewidth}
		\includegraphics[width=\textwidth]{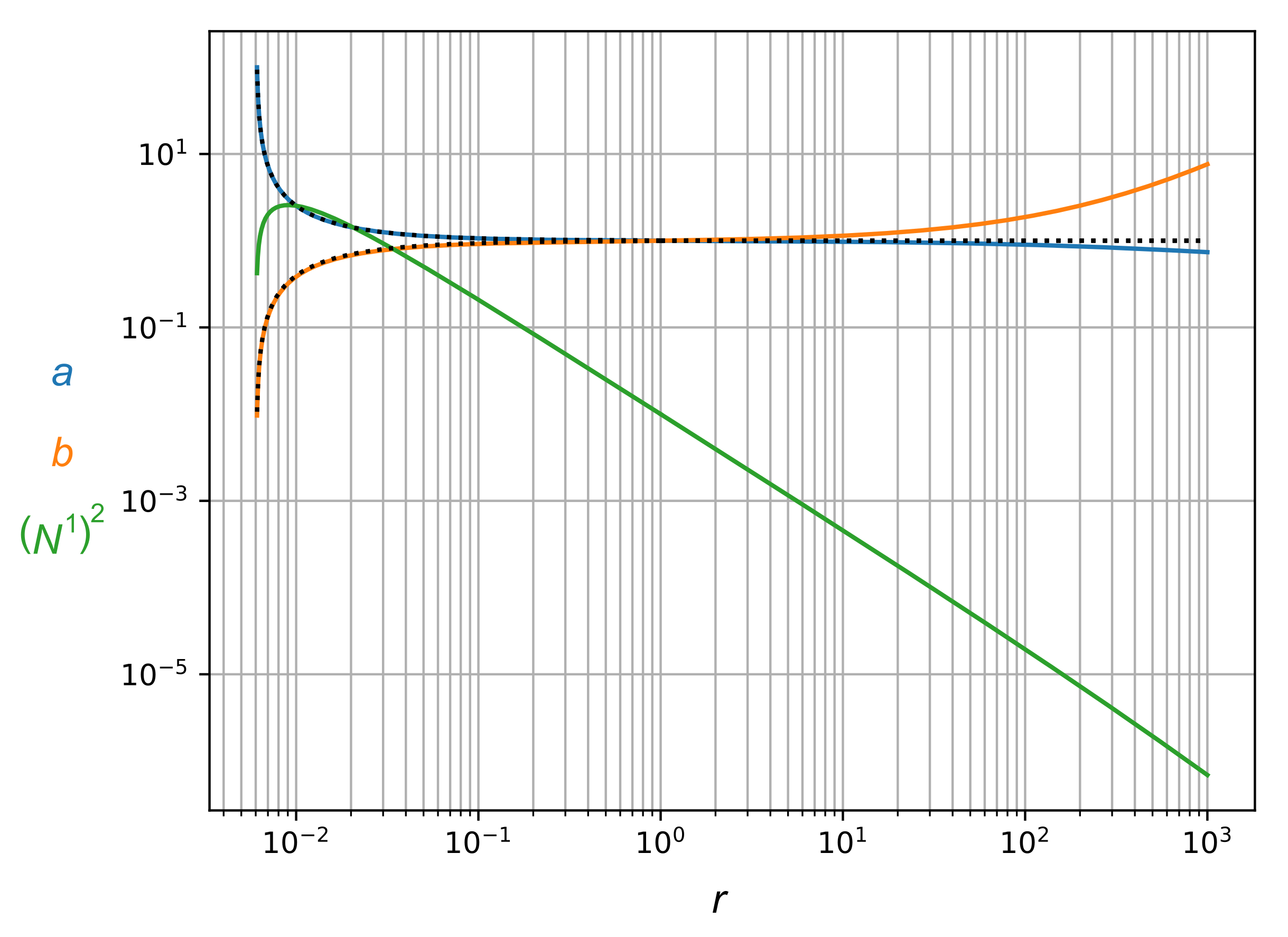}
		\caption{Graph with $w(1)=0.01$ and $a(1)=b(1)=1$. Comparing this with (b) and, taking note of the different vertical scales, one can see the approach
                  to the usual Schwarzschild solution as $w(1)\to 0$. }
	\end{subfigure}
	\hskip2em
	\begin{subfigure}{.45\linewidth}
		\includegraphics[width=\textwidth]{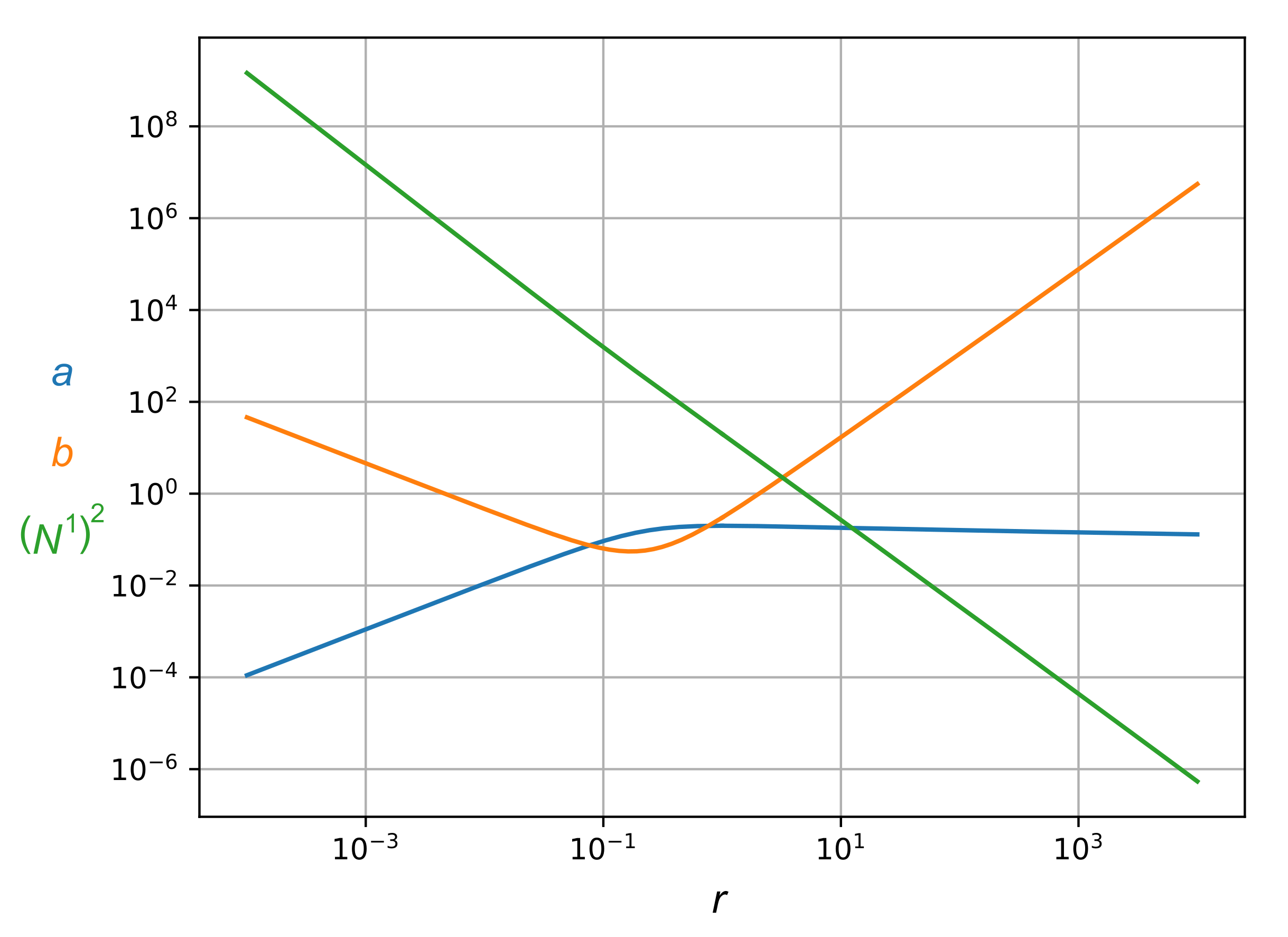}
		\caption{Graph with $w(1)=0.8$, $a(1)=0.2$, and  $b(1)=0.3$ showing a different type of solution with no critical ``black-hole'' radius, but rather a singularity at $r=0$.
                The solution for $b(r)$ still clearly blows up as $r\to\infty$.}
	\end{subfigure}
	\caption{Numerical Solutions of equations \eq{5.22} and \eq{5.23}.}
	\labfig{Sch}
      \end{figure}

  \section{Homogeneous Expanding Universe}
  \setcounter{equation}{0}
  \labsect{heu}

It should be emphasized that the solution given here, which is incompatible with observations, is for a homogeneous universe.
It does not apply to a universe where space-time itself has fluctuations that are not due to ordinary matter. Our theory
implies such fluctuations occur and so one should expect that its cosmological predictions deviate from those
presented in this section. This is explained further in the next section.

We take the Robertson-Walker metric,
\beq ds^2=A^2\left[\frac{d\Gs^2}{1-k\Gs^2}+\Gs^2(d\Gt)^2+\sin^2\Gt\,d\Gf^2)\right]-dt^2 \eeq{6.1}
    where $\Gs=r/A$ and $A$ can be a function of time. With $x_1=\Gs$, $x_2=\Gt$, $x_3=\Gf$, and $x_4=t$ the 
    corresponding metric coefficients are
    \beq g_{11}=A^2/(1-k\Gs^2),\quad g_{22}=A^2\Gs^2,\quad g_{33}=A^2\Gs^2\sin^2\Gt,\quad g_{44}=-1.
    \eeq{6.2}
    Assuming $N^1=N^2=N^3=0$ and defining
    \beq P= 2k+(A\ddot{A}+2\dot{A}^2),
     \eeq{6.3}
where the dot and double dot denote first and second derivatives with respect to time, the equations become
\beqa 0& = & \overline{R}_{11}=P/(1-k\Gs^2)-2(N^4)^2A^2/(1-k\Gs^2) ,\nonum
0 & = &\overline{R}_{22}= P\Gs^2-2(N^4)^2A^2\Gs^2  ,\nonum
0 & = & \overline{R}_{33}=P\Gs^2\sin^2\Gt-2(N^4)^2A^2\Gs^2\sin^2\Gt ,\nonum
0 & = & \overline{R}_{44}=-3\ddot{A}/A,
\eeqa{6.4}
where the terms not involving $N^4$ can be identified with the standard formulas for $R^0_{ij}$. The last equation in \eq{6.3}
implies $\dot{A}$ is a constant that we define to be $\Gb$. We obtain
\beq P=2k+2\dot{A}^2=2(k+\Gb^2), \quad A=\Gb t+ \Gg,
\eeq{6.5}
where $\Gg$ is an integration constant, that we can choose to be zero by redefining our origin of time appropriately.  From the remaining three equations in \eq{6.3}, which are all equivalent, we obtain
\beq (N^4)^2=\frac{P}{2A^2}=\frac{k+\Gb^2}{\Gb^2 t^2},
\eeq{6.6}
which requires that $k\geq -\Gb^2$.

\section{Addressing the dark matter and dark energy problem}
\setcounter{equation}{0}
The result of the previous section giving an expansion rate $\dot{A}$ independent of time agrees with the
well known result that $\ddot{A}=0$ for a model with $p=-\Gm_0/3$. However, this is based on the premise that
spacetime is homogeneous. The expansion of the universe appears to be accelerating with measurements
indicating $p=-0.8\Gm_0$ \cite{Abbott:2019:CCM}, and this could be a consequence of our theory as we now
explain.

Dark matter itself
is known to be inhomogeneous: see, for example, \cite{Nierenberg:2019:DDM} and references therein.
So too is spacetime inhomogeneous in our model. If there is a small fluctuation in the torsion
vector field such that, for example, there is a higher equivalent mass density in two different regions, then there will be
gravitational attraction between these regions, leading to accretion. At the same time ``collisions'' between accreting regions
should tend to disperse the torsion vector field. Thus there will be a certain amount of equivalent kinetic energy associated
with the torsion field accounting for some additional ``dark energy''. More importantly, there could be substructures
in the torsion field containing differing ratios of ``dark energy'' to ``dark mass''.  The structures could collide and give rise to
different structures. In particular, there might be
``negative mass structures'' by  which we mean structures in the torsion vector field incorporating superluminal regions.
Accounting for these effects should reduce the total mass density, providing a higher $p/\Gm_0$ ratio that may
be consistent with the experimental value of $-0.8$.

It is to be emphasized that both our full equations \eq{3.7c}
and their weak field approximations \eq{3.8e}, \eq{3.8f} and \eq{4.b} have no intrinsic length scale. There is a length scale associated
with the overall density of the torsion vector field (connected with the mass density of the apparent dark matter and
dark energy in our theory), but this is of the order of the radius of the universe.  It seems likely
that the  torsion vector field could be quite turbulent with structures on many length scales, down to
some lower cutoff length scale where the current theory breaks down. This cutoff could be the Planck length scale.

To provide quantitative predictions one needs a better idea of the behavior of the torsion vector field within
spacetime, and this will almost certainly require sophisticated numerical simulations to obtain an approximation
to the ``macroscopic equation of state''.
Simulations are needed to provide a better understanding of torsion fluid
behavior in intergalactic and interstellar regions as well as around stars, globular clusters, galaxies, and galaxy clusters. These may require the introduction
of some parameter that provides a lower length scale to the ``turbulence'' in the torsion vector field, that ultimately
could be taken to zero. Simulating the dynamics of the torsion vector field over the continuum of length scales may also require a sort of
numerical renormalization group approach. While we have not investigated the stability of the
torsion waves and torsion rolls, it is not important that they are stable, even in the weak field approximation. The
purpose of our exact solutions in the weak field approximation was mainly to illustrate the rich dynamics of the torsion vector
field and to give some insight into possible dynamics.

Regarding the question as to whether our model can account for the galactic dark mass problem, an encouraging sign is the
apparent cosmological connection between the  critical acceleration $a_0\approx 1.2\times 10^{-10} m/s^2$ in MOND,
the radius of the universe, and the density of dark matter or energy in the universe, as reviewed in \cite{Milgrom:2020:CCM}.
Thus, the density of dark matter or energy, roughly $\Gvr\approx 10^{-27}kg/m^3$, which in our theory is related to the strength of the torsion
field $\BN$, has an associated length scale $1/\sqrt{c^2\Gvr\Gk}\approx 6\times 10^{26}$ meters (approximately the radius of the universe)
which agrees with the length scale $c^2/a_0\approx 7.5\times 10^{26}$ meters associated with the critical acceleration $a_0$ in MOND. 
\section{Conclusion}
\setcounter{equation}{0}

The theory presented here is largely aimed at providing equations governing the behavior of space-time and the torsion field in regions devoid of
matter. An initial test of the theory would entail numerical simulations of a universe where ordinary matter is absent,
allowing for fluctuations in the torsion field. One could start with a homogeneous universe, with the only non-zero component
of $\BN$ being $N_4$ in the metric \eq{6.1}, then add a small spatially random component to $\BN$ and follow the dynamics of $\BN$ along with that of space-time.
Fluctuations in the torsion field should be truncated at a small length scale, perhaps at the Planck length scale.
For the theory to be viable, without modification, the results need to be consistent with cosmological observations.

Beyond the need for a lower cutoff, the equations are still incomplete. As remarked already, one can change the sign of $\BN(\Bx)$ in any region
and still satisfy the equations, indicating that there is a deeper theory that prevents such discontinuous solutions for $\BN(\Bx)$.
Perhaps this also enters at the Planck length scale, and both it and the truncation of fluctuations in the torsion field are accounted for by
appropriate quantum equations. Assuming there is only weak coupling between the torsion fluid with matter,
aside from the coupling due to gravitation (spacetime curvature) then one might think there is conservation of momentum and energy both for
the stress-energy-momentum tensor of the torsion vector field, and for the stress-energy-momentum tensor of matter. On the other hand, if
one regards the conservation of momentum and energy as a consequence of the Bianchi identities then there appears to be no reason why they
should be separately conserved. For this reason our current theory, while it describes the curvature of spacetime and
the accompanying torsion vector field in regions devoid of matter, is incomplete in regions containing matter. 

One appealing feature of Cartan's equations, and which is absent in our current theory, is that they allow for the incorporation of intrinsic spin - something that was
discovered in 1925-26 after Cartan first arrived at his remarkable equations. Cartan was originally motivated by the work of the
Cosserat brothers \cite{Cosserat:1968:FPC} which, like his
equations, allowed for a non-symmetric stress field. His focus was on deriving equations where the source (matter) field
automatically satisfied energy and momentum conservation. Sciama \cite{Sciama:1962:RDG} and Kibble \cite{Kibble:1961:SBN} independently developed the same
generalization of Cartan's theory, known as $U_4$ or the Einstein-Cartan-(Sciama-Kibble) theory. Their theory and the original Cartan theory reduces to the Einstein equations when matter is not present. An additional advantage of these theories, not yet incorporated in our theory as there is no
coupling with matter, is that they account for conservation of angular momentum \cite{Hehl:1976:GRS}.

As others have also realized, departing from Cartan's approach
has the potential for explaining dark energy and dark matter as manifestations of a revised gravitational theory.
Our theory is perhaps the simplest theory with that potential. As stressed already, conservation of energy and momentum still hold
provided one reinterprets the equations as Einstein's equation with an energy-momentum-stress tensor associated with "empty space",
i.e. associated with the torsion field. It could be that more complicated equations involving torsion  will provide the final answer
(and, as observed in the introduction, many candidates, besides Cartan's and those of Sciama and Kibble, have been proposed, and undoubtedly others will be put forward in the future).
In that case, it could be that the ultimate theory only slightly perturbs the results in our theory in the intergalactic and interstellar regions,
yet provides some lower limit to the likely "turbulence" in the torsion field. Thus, if successful, the theory proposed here may provide a guide in the search
for the ultimate theory. It may be that the most important ``take home'' message of this paper is highlighting the
importance of considering torsion theories that allow for dynamics in empty space
on multiple length scales of the torsion field (and hence of the accompanying metric).

If warranted by experimental observations, a natural modification of our theory would be to add a term involving
Einstein's cosmological constant $\GL$. But it would be far more satisfying if this was not needed.

\section*{Acknowledgments}
The author thanks the Whitlam Government of Australia for providing support in the late 1970's through free university education and a stipend while
the author was an undergraduate, which was when this work was initiated. Additionally, the author thanks Sydney University, Cornell University,
Caltech, the Courant Institute and the University of Utah, and the National Science Foundation, through a succession of grants, for support.
Christian Kern is thanked for his help with the figures, for the numerical simulations needed to produce them, and in particular for his
discovery of the example in \fig{Sch}(d). M. Milgrom is thanked for helpful
comments, recently and dating back to the early 1990's, for supplying Figure 1, noticing some minor errors, and
providing many useful references.

\ifx \bblindex \undefined \def \bblindex #1{} \fi\ifx \bbljournal \undefined
  \def \bbljournal #1{{\em #1}\index{#1@{\em #1}}} \fi\ifx \bblnumber
  \undefined \def \bblnumber #1{{\bf #1}} \fi\ifx \bblvolume \undefined \def
  \bblvolume #1{{\bf #1}} \fi\ifx \noopsort \undefined \def \noopsort #1{}
  \fi\ifx \bblindex \undefined \def \bblindex #1{} \fi\ifx \bbljournal
  \undefined \def \bbljournal #1{{\em #1}\index{#1@{\em #1}}} \fi\ifx
  \bblnumber \undefined \def \bblnumber #1{{\bf #1}} \fi\ifx \bblvolume
  \undefined \def \bblvolume #1{{\bf #1}} \fi\ifx \noopsort \undefined \def
  \noopsort #1{} \fi


\end{document}